\documentclass[submission]{eptcs}
\providecommand{\event}{Transformation Tool Contest 2013 (TTC'13)}
\def\titlerunning{Solving the TTC 2013 Flowgraphs Case with FunnyQT}
\def\authorrunning{Tassilo Horn}

\usepackage[utf8]{inputenc}
\usepackage[T1]{fontenc}
\usepackage{hyperref}
\usepackage{paralist}
\usepackage{fancyvrb}

\usepackage{color}
\input{pyg}

\makeatletter
\def\verbatim@font{\ttfamily\small}
\makeatother

\title{Solving the TTC 2013 Flowgraphs Case with FunnyQT}
\author{Tassilo Horn
  \email{horn@uni-koblenz.de}
  \institute{Institute for Software Technology, University Koblenz-Landau, Germany}}

\begin{document}

\maketitle

\begin{abstract}
  FunnyQT is a model querying and model transformation library for the
  functional Lisp-dialect Clojure providing a rich and efficient querying and
  transformation API.

  This paper describes the FunnyQT solution to the TTC 2013 Flowgraphs
  Transformation Case.  It solves all four tasks, and it has won the \emph{best
    efficiency award} for this case.
\end{abstract}

\section{Introduction}
\label{sec:introduction}

\emph{FunnyQT}\footnote{The FunnyQT homepage:
  \url{https://github.com/jgralab/funnyqt}} is a new model querying and
transformation approach which is implemented as an API for the functional,
JVM-based Lisp-dialect Clojure.  It provides several sub-APIs for implementing
different kinds of queries and transformations.  For example, there is a
model-to-model transformation API, and there is an in-place transformation API
for writing programmed graph transformations.  FunnyQT currently works natively
with EMF and JGraLab models, i.e., it doesn't use some internal model
representation, and it can be extended to other modeling frameworks, too.

For solving the tasks of this transformation case\footnote{This FunnyQT
  solution is available at \url{https://github.com/tsdh/ttc-2013-flowgraphs}
  and on SHARE (image \textsf{TTC13::Ubuntu12LTS\_TTC13::FunnyQT.vdi}},
FunnyQT's model transformation API and its polymorphic function API have been
used for task~1.  Both task~2 and task~3 have been tackled algorithmically
using FunnyQT's plain querying and model manipulation APIs.  Task~4 has been
solved by using FunnyQT's querying API and Clojure metaprogramming.

\section{Solution Description}
\label{sec:solution-description}

\paragraph{Task 1: JaMoPP to StructureGraph.}
\label{sec:task-1}

According to the case description \cite{flowgraphcasedesc}, the goal of this
task is to transform a fine-granular Java syntax graph conforming to the JaMoPP
metamodel \cite{jamopp09} into a much simpler structure graph model that only
contains statements and expressions that are neither structured nor subdivided
any further.  However, the original Java code of these statements and
expressions should be reflected in the new elements' \verb|txt| attribute.
This model-to-text transformation is described in the next paragraph.
Thereafter, the model-to-model transformation creating a structure graph from a
JaMoPP model is described.

\subparagraph{JaMoPP to Text.}
\label{sec:jamopp-text}

This model-to-text transformation is implemented using FunnyQT's polymorphic
function API.  A polymorphic function is a function that is declared once, and
then arbitrarily many implementations for concrete metamodel types can be
added.  When a polymorphic function is called, the actual implementation is
determined similarly to the typical dispatch in object-oriented programming
languages.  If there is no implementation provided for the element's type or
one of its supertypes, an exception is thrown.

The function \verb|stmt2str| implements the model-to-text transformation
required for solving task~1.  It is declared as follows.

\begin{Verbatim}[commandchars=\\\{\},fontsize=\footnotesize]
\PY{p}{(}\PY{n+nf}{declare\PYZhy{}polyfn} \PY{n+nv}{stmt2str} \PY{p}{[}\PY{n+nv}{elem}\PY{p}{]}\PY{p}{)}
\end{Verbatim}

\verb|declare-polyfn| declares a new polymorphic function.  Its name is
\verb|stmt2str|, and it receives exactly one parameter \verb|elem|.  Its task
is to create a string representation matching the concrete Java syntax for the
provided JaMoPP model element.

After the polymorphic function has been declared, implementations for concrete
metamodel types can be added using \verb|defpolyfn|.  For example, this is the
implementation for JaMoPP elements of type \verb|AssignmentExpression|:

\begin{Verbatim}[commandchars=\\\{\},fontsize=\footnotesize]
\PY{p}{(}\PY{n+nf}{defpolyfn} \PY{n+nv}{stmt2str} \PY{l+s+ss}{\PYZsq{}expressions.AssignmentExpression} \PY{p}{[}\PY{n+nv}{ae}\PY{p}{]}
  \PY{p}{(}\PY{n+nb}{str }\PY{p}{(}\PY{n+nf}{stmt2str} \PY{p}{(}\PY{n+nf}{eget} \PY{n+nv}{ae} \PY{l+s+ss}{:child}\PY{p}{)}\PY{p}{)} \PY{l+s}{\PYZdq{} \PYZdq{}}
       \PY{p}{(}\PY{n+nf}{stmt2str} \PY{p}{(}\PY{n+nf}{eget} \PY{n+nv}{ae} \PY{l+s+ss}{:assignmentOperator}\PY{p}{)}\PY{p}{)} \PY{l+s}{\PYZdq{} \PYZdq{}}
       \PY{p}{(}\PY{n+nf}{stmt2str} \PY{p}{(}\PY{n+nf}{eget} \PY{n+nv}{ae} \PY{l+s+ss}{:value}\PY{p}{)}\PY{p}{)}\PY{p}{)}\PY{p}{)}
\end{Verbatim}

The \verb|child| of the assignment expression is some variable, the
\verb|assignmentOperator| is one of \verb|=|, \verb|+=|, \verb|-=|, \verb|*=|,
or \verb|/=|, and \verb|value| is an arbitrary expression.  These three
components are converted to strings using \verb|stmt2str| again which are then
concatenated.

All in all, the polymorphic \verb|stmt2str| function consists of 22
implementations for various JaMoPP metamodel types accounting to a total of 107
lines of code.  The complete model-to-text transformation is printed in
Appendix~\ref{sec:complete-jamopp-text}.

\subparagraph{JaMoPP to Structure Graph.}
\label{sec:jamopp-struct-graph}

The JaMoPP-to-StructureGraph transformation is implemented using FunnyQT's
model-to-model transformation API.  This transformation also creates \verb|Var|
and \verb|Param| objects as requested by task~3.1.

The transformation starts by defining its name and input and output models.

\begin{Verbatim}[commandchars=\\\{\},fontsize=\footnotesize]
\PY{p}{(}\PY{n+nf}{deftransformation} \PY{n+nv}{java2flowgraph} \PY{p}{[}\PY{p}{[}\PY{n+nv}{in} \PY{l+s+ss}{:emf}\PY{p}{]} \PY{p}{[}\PY{n+nv}{out} \PY{l+s+ss}{:emf}\PY{p}{]}\PY{p}{]}
\end{Verbatim}

There could be arbitrarily many input and output models, and they could be of
different kinds, e.g., a transformation could receive a JGraLab TGraph and some
EMF model, and create an output EMF model.  Here, it gets only the JaMoPP EMF
input model which is bound to the variable \verb|in|, and one single structure
graph output model bound to \verb|out|, which is also an EMF model.

In the body of such a transformation, arbitrarily many rules may be defined.
The first one is the \verb|method2method| rule shown in the next listing.  The
\verb|^:top| metadata preceeding the rule name specifies that the rule is a
top-level rule.  Such rules are applied to all matching elements by the
transformation itself, whereas non-top-level rules have to be called explicitly
from a top-level rule (directly or indirectly).

\begin{Verbatim}[commandchars=\\\{\},fontsize=\footnotesize]
  \PY{p}{(}\PY{o}{\PYZca{}}\PY{l+s+ss}{:top} \PY{n+nv}{method2method} \PY{p}{[}\PY{n+nv}{m}\PY{p}{]}
         \PY{l+s+ss}{:from} \PY{l+s+ss}{\PYZsq{}members.ClassMethod}
         \PY{l+s+ss}{:to} \PY{p}{[}\PY{n+nv}{fgm} \PY{l+s+ss}{\PYZsq{}flowgraph.Method}, \PY{n+nv}{fgex} \PY{l+s+ss}{\PYZsq{}flowgraph.Exit}\PY{p}{]}
         \PY{p}{(}\PY{n+nf}{eset!} \PY{n+nv}{fgm} \PY{l+s+ss}{:txt} \PY{p}{(}\PY{n+nf}{stmt2str} \PY{n+nv}{m}\PY{p}{)}\PY{p}{)} \PY{c+c1}{;; Invoke the model\PYZhy{}to\PYZhy{}text transformation}
         \PY{p}{(}\PY{n+nf}{eset!} \PY{n+nv}{fgex} \PY{l+s+ss}{:txt} \PY{l+s}{\PYZdq{}Exit\PYZdq{}}\PY{p}{)}
         \PY{p}{(}\PY{n+nf}{eset!} \PY{n+nv}{fgm} \PY{l+s+ss}{:exit} \PY{n+nv}{fgex}\PY{p}{)}
         \PY{p}{(}\PY{n+nf}{eset!} \PY{n+nv}{fgm} \PY{l+s+ss}{:stmts} \PY{p}{(}\PY{n+nb}{map }\PY{n+nv}{stmt2item} \PY{p}{(}\PY{n+nf}{eget} \PY{n+nv}{m} \PY{l+s+ss}{:statements}\PY{p}{)}\PY{p}{)}\PY{p}{)}  \PY{c+c1}{;; transform the statements}
         \PY{p}{(}\PY{n+nf}{eset!} \PY{n+nv}{fgm} \PY{l+s+ss}{:def} \PY{p}{(}\PY{n+nb}{map }\PY{n+nv}{param2param} \PY{p}{(}\PY{n+nf}{eget} \PY{n+nv}{m} \PY{l+s+ss}{:parameters}\PY{p}{)}\PY{p}{)}\PY{p}{)}\PY{p}{)} \PY{c+c1}{;; transform the parameters}
\end{Verbatim}

The rule receives a JaMoPP model element \verb|m|.  The \verb|:from| clause
dictates that \verb|m| must be of type \verb|ClassMethod| in order for the rule
to be applicable.  The \verb|:to| clause declares the objects to be created.
Here, for a given JaMoPP method, a corresponding flowgraph method and its exit
object are created.  The remainder of the rule is its body containing arbitrary
code to set attributes and references.  Here, the \verb|txt| attribute of the
new method and its exit are set, the former using the polymorphic
\verb|stmt2str| function discussed above.  The method's \verb|stmts| reference
is set by applying another rule, \verb|stmt2item|, to the statements of the
JaMoPP method.  Likewise, the method's parameters are transformed by mapping
them to the \verb|param2param| rule for setting the method's \verb|def|
reference.

Special kinds of rules are generalizing rules such as the one shown in the next
listing.

\begin{Verbatim}[commandchars=\\\{\},fontsize=\footnotesize]
  \PY{p}{(}\PY{n+nf}{stmt2item} \PY{p}{[}\PY{n+nv}{stmt}\PY{p}{]}
      \PY{l+s+ss}{:generalizes} \PY{p}{[}\PY{n+nv}{local\PYZhy{}var\PYZhy{}stmt2simple\PYZhy{}stmt} \PY{n+nv}{condition2if} \PY{n+nv}{block2block}
                    \PY{n+nv}{return2return} \PY{n+nv}{while\PYZhy{}loop2loop} \PY{n+nv}{break2break} \PY{n+nv}{continue2continue}
                    \PY{n+nv}{label2label} \PY{n+nv}{stmt2simple\PYZhy{}stmt}\PY{p}{]}\PY{p}{)}
\end{Verbatim}

This concept is quite similar to mapping disjunction in QVT Operational
Mappings.  When this rule is called, the rules specified in the
\verb|:generalizes| vector are tried one after the other, and the first
applicable one is applied, and its result is returned.  Furthermore, a
generalizing rule also combines the traceability mappings of all subrules.

The complete \verb|java2flowgraph| model-to-model transformation consists of 15
rules with 93 lines of code in total.  It is printed in
Appendix~\ref{sec:compl-jamopp-struct}.

\paragraph{Task 2: Control Flow Analysis.}
\label{sec:task-2}

The purpose of this task is to create \verb|cfNext| links between
\verb|FlowInstr| elements in the flowgraph model created by the model-to-model
transformation realizing task~1.  Every such flow instruction should be
connected to every other flow instruction that may be the next one in the
program's control flow.  This challenge has been tackled algorithmically using
FunnyQT's plain quering and model manipulation APIs.

The algorithm uses a sequence of statements as intermediate representation to
work on realizing a pre-order depth-first traversal with look-ahead through the
method's statements.  In the general case, every flow instruction in that
sequence has to be connected with the immediately following flow instruction in
the sequence.  For various kinds of statements, special rules are needed.  For
example, when encountering a block in the sequence (which is no flow
instruction), the block is replaced with its contents.

Since the next statement in the sequence might not be a flow instruction but
some structured statement like a block, an if-statement, or a loop, there is a
helper function \verb|cf-peek|.  It receives some element and returns either
this element if it is a flow instruction, or otherwise the first flow
instruction inside this element.

The function \verb|cf-synth| synthesizing the control flow links using the
algorithm sketched above is explained in the next four listings.  It receives
the sequence of statements \verb|v|, the method's \verb|Exit| node \verb|exit|,
the current loop's test expression (\verb|loop-expr|), the statement following
the current loop (\verb|loop-succ|), and a map \verb|label-succ-map| that
assigns to each label reachable in the current scope the statement following
the labeled statement.  The \verb|exit| parameter is used for handling return
statements, and the last three parameters are used for handling break and
continue statements.  Initially, the function is called with \verb|v| only
containing the method, and \verb|exit| bound to that method's exit.  All other
parameters are \verb|nil|.

\begin{Verbatim}[commandchars=\\\{\},fontsize=\footnotesize]
\PY{p}{(}\PY{k+kd}{defn }\PY{n+nv}{cf\PYZhy{}synth} \PY{p}{[}\PY{n+nv}{v} \PY{n+nv}{exit} \PY{n+nv}{loop\PYZhy{}expr} \PY{n+nv}{loop\PYZhy{}succ} \PY{n+nv}{label\PYZhy{}succ\PYZhy{}map}\PY{p}{]}
  \PY{p}{(}\PY{n+nb}{when }\PY{p}{(}\PY{n+nb}{seq }\PY{n+nv}{v}\PY{p}{)}
    \PY{p}{(}\PY{k}{let }\PY{p}{[}\PY{p}{[}\PY{n+nv}{el} \PY{o}{\PYZam{}} \PY{p}{[}\PY{n+nv}{n} \PY{o}{\PYZam{}} \PY{n+nv}{\PYZus{}} \PY{l+s+ss}{:as} \PY{n+nv}{tail}\PY{p}{]}\PY{p}{]} \PY{n+nv}{v}\PY{p}{]}
      \PY{p}{(}\PY{n+nf}{type\PYZhy{}case} \PY{n+nv}{el}
\end{Verbatim}

If the sequence \verb|v| is not empty, its first element is bound to \verb|el|,
and its rest is bound to \verb|tail|.  Furthermore, the first element of the
rest (i.e., the second element of the sequence) is bound to \verb|n|.

After binding these elements, a \verb|type-case| dispatches on \verb|el|'s
metamodel type.  For example, if the element is a method, a control flow link
to that method's first flow instruction is created, and the function recurses
with the method's statements (\verb|recur| is an explicit tail-recursive call).

\begin{Verbatim}[commandchars=\\\{\},fontsize=\footnotesize]
        \PY{l+s+ss}{\PYZsq{}flowgraph.Method} \PY{p}{(}\PY{k}{let }\PY{p}{[}\PY{n+nv}{stmts} \PY{p}{(}\PY{n+nf}{econtents} \PY{n+nv}{el}\PY{p}{)}\PY{p}{]}
                            \PY{p}{(}\PY{n+nf}{eadd!} \PY{n+nv}{el} \PY{l+s+ss}{:cfNext} \PY{p}{(}\PY{n+nf}{cf\PYZhy{}peek} \PY{p}{(}\PY{n+nb}{first }\PY{n+nv}{stmts}\PY{p}{)}\PY{p}{)}\PY{p}{)}
                            \PY{p}{(}\PY{n+nf}{recur} \PY{n+nv}{stmts} \PY{n+nv}{exit} \PY{n+nv}{nil} \PY{n+nv}{nil} \PY{n+nv}{nil}\PY{p}{)}\PY{p}{)}
\end{Verbatim}

If the current element is a label, the function recurses with that label's
statement prepended to the tail of the sequence.  A mapping from this label to
its following statement is added to the \verb|label-succ-map|.  This
statement's first flow instruction is where the control flow continues when
breaking to this label.

\begin{Verbatim}[commandchars=\\\{\},fontsize=\footnotesize]
        \PY{l+s+ss}{\PYZsq{}flowgraph.Label} \PY{p}{(}\PY{n+nf}{recur} \PY{p}{(}\PY{n+nb}{cons }\PY{p}{(}\PY{n+nf}{eget} \PY{n+nv}{el} \PY{l+s+ss}{:stmt}\PY{p}{)} \PY{n+nv}{tail}\PY{p}{)} \PY{n+nv}{exit} \PY{n+nv}{loop\PYZhy{}expr} \PY{n+nv}{loop\PYZhy{}succ}
                                \PY{p}{(}\PY{n+nb}{assoc }\PY{n+nv}{label\PYZhy{}succ\PYZhy{}map} \PY{n+nv}{el} \PY{n+nv}{n}\PY{p}{)}\PY{p}{)}
\end{Verbatim}

If the current element is a break statement, two cases have to be
distinguished.  If the break is labeled, a control flow link is added to the
first flow instruction of the statement following the label which can be looked
up in the \verb|label-succ-map|.  If the break is not labled, a control flow
link is added to the first flow instruction in the statement following the
surrounding loop which is bound to \verb|loop-succ|.

In any case, the function recurses with the tail of the sequence keeping all
other parameters as-is.

\begin{Verbatim}[commandchars=\\\{\},fontsize=\footnotesize]
        \PY{l+s+ss}{\PYZsq{}flowgraph.Break} \PY{p}{(}\PY{k}{do }\PY{p}{(}\PY{n+nb}{if\PYZhy{}let }\PY{p}{[}\PY{n+nv}{l} \PY{p}{(}\PY{n+nf}{eget} \PY{n+nv}{el} \PY{l+s+ss}{:label}\PY{p}{)}\PY{p}{]}
                               \PY{p}{(}\PY{n+nf}{eadd!} \PY{n+nv}{el} \PY{l+s+ss}{:cfNext} \PY{p}{(}\PY{n+nf}{cf\PYZhy{}peek} \PY{p}{(}\PY{n+nf}{label\PYZhy{}succ\PYZhy{}map} \PY{n+nv}{l}\PY{p}{)}\PY{p}{)}\PY{p}{)}
                               \PY{p}{(}\PY{n+nf}{eadd!} \PY{n+nv}{el} \PY{l+s+ss}{:cfNext} \PY{p}{(}\PY{n+nf}{cf\PYZhy{}peek} \PY{n+nv}{loop\PYZhy{}succ}\PY{p}{)}\PY{p}{)}\PY{p}{)}
                           \PY{p}{(}\PY{n+nf}{recur} \PY{n+nv}{tail} \PY{n+nv}{exit} \PY{n+nv}{loop\PYZhy{}expr} \PY{n+nv}{loop\PYZhy{}succ} \PY{n+nv}{label\PYZhy{}succ\PYZhy{}map}\PY{p}{)}\PY{p}{)}
\end{Verbatim}

There are similar cases for handling objects of the other metamodel types.  The
complete control flow transformation consists of 57 lines of code and is
printed in Appendix~\ref{sec:compl-contr-flow}.

\paragraph{Task 3: Data Flow Analysis.}
\label{sec:task-3}

The purpose of this task is to create \verb|dfNext| links between
\verb|FlowInst| elements where the target element is a control flow successor
of the source element, the target element uses (reads) a variable that was
defined (written) by the source element, and the variable has not been
rewritten in between.  This definition has been implemented exactly as stated
here, because although it's not the most efficient algorithm for the task, it
is very clear and concise.

The function \verb|find-nearest-definers| receives a flow instruction \verb|fi|
and a variable \verb|uv| used by it, and it returns a vector of the nearest
control flow predecessors that define that variable.

\begin{Verbatim}[commandchars=\\\{\},fontsize=\footnotesize]
\PY{p}{(}\PY{k+kd}{defn }\PY{n+nv}{find\PYZhy{}nearest\PYZhy{}definers} \PY{p}{[}\PY{n+nv}{fi} \PY{n+nv}{uv}\PY{p}{]}
  \PY{p}{(}\PY{k}{loop }\PY{p}{[}\PY{n+nv}{preds} \PY{p}{(}\PY{n+nb}{mapcat }\PY{o}{\PYZsh{}}\PY{p}{(}\PY{n+nf}{adjs} \PY{n+nv}{\PYZpc{}} \PY{l+s+ss}{:cfPrev}\PY{p}{)} \PY{p}{(}\PY{k}{if }\PY{p}{(}\PY{n+nf}{coll?} \PY{n+nv}{fi}\PY{p}{)} \PY{n+nv}{fi} \PY{p}{[}\PY{n+nv}{fi}\PY{p}{]}\PY{p}{)}\PY{p}{)},
         \PY{n+nv}{r} \PY{p}{[}\PY{p}{]}, \PY{n+nv}{known} \PY{o}{\PYZsh{}}\PY{p}{\PYZob{}}\PY{p}{\PYZcb{}}\PY{p}{]}
    \PY{p}{(}\PY{k}{if }\PY{p}{(}\PY{n+nb}{seq }\PY{n+nv}{preds}\PY{p}{)}
      \PY{p}{(}\PY{k}{let }\PY{p}{[}\PY{n+nv}{definers} \PY{p}{(}\PY{n+nb}{filter }\PY{o}{\PYZsh{}}\PY{p}{(}\PY{n+nf}{member?} \PY{n+nv}{uv} \PY{p}{(}\PY{n+nf}{eget} \PY{n+nv}{\PYZpc{}} \PY{l+s+ss}{:def}\PY{p}{)}\PY{p}{)} \PY{n+nv}{preds}\PY{p}{)}
            \PY{n+nv}{others}   \PY{p}{(}\PY{n+nb}{remove }\PY{o}{\PYZsh{}}\PY{p}{(}\PY{n+nf}{member?} \PY{n+nv}{uv} \PY{p}{(}\PY{n+nf}{eget} \PY{n+nv}{\PYZpc{}} \PY{l+s+ss}{:def}\PY{p}{)}\PY{p}{)} \PY{n+nv}{preds}\PY{p}{)}\PY{p}{]}
        \PY{p}{(}\PY{n+nf}{recur} \PY{p}{(}\PY{n+nb}{remove }\PY{o}{\PYZsh{}}\PY{p}{(}\PY{n+nf}{member?} \PY{n+nv}{\PYZpc{}} \PY{n+nv}{known}\PY{p}{)} \PY{p}{(}\PY{n+nb}{mapcat }\PY{o}{\PYZsh{}}\PY{p}{(}\PY{n+nf}{adjs} \PY{n+nv}{\PYZpc{}} \PY{l+s+ss}{:cfPrev}\PY{p}{)} \PY{n+nv}{others}\PY{p}{)}\PY{p}{)}
               \PY{p}{(}\PY{n+nb}{into }\PY{n+nv}{r} \PY{n+nv}{definers}\PY{p}{)} \PY{p}{(}\PY{n+nb}{into }\PY{n+nv}{known} \PY{n+nv}{preds}\PY{p}{)}\PY{p}{)}\PY{p}{)}
      \PY{n+nv}{r}\PY{p}{)}\PY{p}{)}\PY{p}{)}
\end{Verbatim}

In Clojure, \verb|loop| and \verb|recur| implement a local tail-recursion, that
is, inside a \verb|loop| a \verb|recur| form recurses not to the surrounding
function but to the surrounding \verb|loop|.  Initially, \verb|preds| is bound
to the immediate control flow predecessors of \verb|fi|, the result variable
\verb|r| is bound to the empty vector, and \verb|known| is bound to the empty
set.

If there are no predecessors, the result \verb|r| is returned (the else-branch
of the if).  If there are control flow predecessors, those are sorted into
\verb|definers| and \verb|others|, i.e., flow instructions that write to
\verb|uv|, and flow instructions that do not write to \verb|uv|, respectively.

Then it is recursed to the surrounding \verb|loop|.  \verb|preds| is rebound to
those control flow predecessors of \verb|others| that aren't already known in
order not to recurse infinitely in case of control flow cycles, the result
vector \verb|r| is rebound to the current \verb|r| value plus the new
\verb|definers|, and \verb|known| is rebound to the union of the current
\verb|known| value and the current \verb|preds|.

The main function of this task simply uses this function to find the nearest
definers of all flow instructions and their used variables and creates
\verb|dfNext| links.

The complete data flow transformation consists of 19 lines of code and is
printed in Appendix~\ref{sec:complete-data-flow}.

\paragraph{Task 4: Control and Data Flow Validation.}
\label{sec:task-4}

The goal of task 4 is to enable offloading testing effort for the
transformations solving tasks 1 to 3 to programmers knowing only Java by
equipping them with some easy to use DSL.  The next listing shows an example
validation specification as provided by the FunnyQT solution.

\begin{Verbatim}[commandchars=\\\{\},fontsize=\footnotesize]
\PY{p}{(}\PY{n+nf}{make\PYZhy{}test} \PY{n+nv}{test\PYZhy{}fg\PYZhy{}transform\PYZhy{}test0} \PY{l+s}{\PYZdq{}models/Test0.java.xmi\PYZdq{}}
           \PY{o}{\PYZsh{}}\PY{p}{\PYZob{}}\PY{p}{[}\PY{l+s}{\PYZdq{}testMethod()\PYZdq{}}   \PY{l+s}{\PYZdq{}int a = 1;\PYZdq{}}\PY{p}{]}     \PY{c+c1}{;; expected cfNext links}
             \PY{c+c1}{;; more [cf\PYZhy{}predecessor cf\PYZhy{}successor] tuples}
             \PY{p}{[}\PY{l+s}{\PYZdq{}return b * c;\PYZdq{}}  \PY{l+s}{\PYZdq{}Exit\PYZdq{}}\PY{p}{]}\PY{p}{\PYZcb{}}
           \PY{o}{\PYZsh{}}\PY{p}{\PYZob{}}\PY{p}{[}\PY{l+s}{\PYZdq{}int a = 1;\PYZdq{}}     \PY{l+s}{\PYZdq{}int c = a + b;\PYZdq{}}\PY{p}{]} \PY{c+c1}{;; expected dfNext links}
             \PY{c+c1}{;; more [df\PYZhy{}predecessor df\PYZhy{}successor] tuples}
             \PY{p}{[}\PY{l+s}{\PYZdq{}b = a \PYZhy{} b;\PYZdq{}}     \PY{l+s}{\PYZdq{}return b * c;\PYZdq{}}\PY{p}{]}\PY{p}{\PYZcb{}}\PY{p}{)}
\end{Verbatim}

The FunnyQT solution uses Clojure's metaprogramming facilities to create an
\emph{internal validation DSL}.  \verb|make-test| is a \emph{macro}.  A macro
is a function that will be called by the Clojure compiler at compile-time.  It
receives the unevaluated arguments given to it, that is, its parameters are
bound to code.  Clojure, like all Lisps, is \emph{homoiconic}, meaning that
code is represented using usual Clojure data structures, e.g., lists, vectors,
symbols, literals, etc.  Thus, the macro is able to transform the code provided
to it using standard Clojure functions to some new bunch of code that takes its
place.  Here, \verb|make-test| creates a unit test that loads the given XMI
model and compares it against the expected control and data flow links.

The complete macro implementation and two complete validation specifications
are printed in Appendix~\ref{sec:compl-valid-dsl}.

\section{Evaluation}
\label{sec:evaluation}

In this section, the FunnyQT solution to the Flowgraphs case is evaluated
according to the critera listed in the case description
\cite{flowgraphcasedesc}.

All four tasks have been solved, and the results of every task are
\emph{complete and correct}.  The FunnyQT solution consists of 313 lines of
code excluding comments and empty lines, making it the shortest of all provided
solutions.  It is also the solution with the best \emph{performance} and has
won the \emph{best efficiency award} for this case.  However, because FunnyQT
is a Clojure API with a functional alignment, its \emph{understandability}
depends largely on a reader's prior knowledge about Clojure and functional
programming.

\bibliographystyle{eptcs}
\bibliography{ttc13-funnyqt-flowgraph}

\begin{thebibliography}{1}
\providecommand{\bibitemdeclare}[2]{}
\providecommand{\surnamestart}{}
\providecommand{\surnameend}{}
\providecommand{\urlprefix}{Available at }
\providecommand{\url}[1]{\texttt{#1}}
\providecommand{\href}[2]{\texttt{#2}}
\providecommand{\urlalt}[2]{\href{#1}{#2}}
\providecommand{\doi}[1]{doi:\urlalt{http://dx.doi.org/#1}{#1}}
\providecommand{\bibinfo}[2]{#2}

\bibitemdeclare{techreport}{jamopp09}
\bibitem{jamopp09}
\bibinfo{author}{Florian \surnamestart Heidenreich\surnameend},
  \bibinfo{author}{Jendrik \surnamestart Johannes\surnameend},
  \bibinfo{author}{Mirko \surnamestart Seifert\surnameend} \&
  \bibinfo{author}{Christian \surnamestart Wende\surnameend}
  (\bibinfo{year}{2009}): \emph{\bibinfo{title}{{JaMoPP: The Java Model Parser
  and Printer}}}.
\newblock \bibinfo{type}{Technical Report} \bibinfo{number}{TUD-FI09-10},
  \bibinfo{institution}{Technische Universität Dresden, Fakult\"at
  Informatik}.
\newblock
  \bibinfo{note}{\url{ftp://ftp.inf.tu-dresden.de/pub/berichte/tud09-10.pdf}}.

\bibitemdeclare{inproceedings}{flowgraphcasedesc}
\bibitem{flowgraphcasedesc}
\bibinfo{author}{Tassilo \surnamestart Horn\surnameend} (\bibinfo{year}{2013}):
  \emph{\bibinfo{title}{The {TTC} 2013 Flowgraphs Case}}.
\newblock In \bibinfo{editor}{Pieter \surnamestart {Van Gorp}\surnameend},
  \bibinfo{editor}{Louis \surnamestart Rose\surnameend} \&
  \bibinfo{editor}{Christian \surnamestart Krause\surnameend}, editors: {\sl
  \bibinfo{booktitle}{Sixth Transformation Tool Contest (TTC 2013)}}, {\sl
  \bibinfo{series}{EPTCS,}} \bibinfo{volume}{this volume}.

\end{thebibliography}

\appendix
\newpage
\section{The complete JaMoPP-to-Text Transformation}
\label{sec:complete-jamopp-text}

\begin{Verbatim}[commandchars=\\\{\},fontsize=\footnotesize]
\PY{p}{(}\PY{n+nf}{declare\PYZhy{}polyfn} \PY{n+nv}{stmt2str} \PY{p}{[}\PY{n+nv}{elem}\PY{p}{]}\PY{p}{)}

\PY{p}{(}\PY{k+kd}{defn }\PY{n+nv}{reduce\PYZhy{}str} \PY{p}{[}\PY{n+nv}{els}\PY{p}{]}
  \PY{p}{(}\PY{n+nb}{reduce }\PY{o}{\PYZsh{}}\PY{p}{(}\PY{n+nb}{str }\PY{n+nv}{\PYZpc{}1} \PY{p}{(}\PY{n+nf}{stmt2str} \PY{n+nv}{\PYZpc{}2}\PY{p}{)}\PY{p}{)} \PY{l+s}{\PYZdq{}\PYZdq{}} \PY{n+nv}{els}\PY{p}{)}\PY{p}{)}

\PY{p}{(}\PY{n+nf}{defpolyfn} \PY{n+nv}{stmt2str} \PY{l+s+ss}{\PYZsq{}members.ClassMethod} \PY{p}{[}\PY{n+nv}{method}\PY{p}{]}
  \PY{p}{(}\PY{n+nb}{str }\PY{p}{(}\PY{n+nf}{eget} \PY{n+nv}{method} \PY{l+s+ss}{:name}\PY{p}{)} \PY{l+s}{\PYZdq{}()\PYZdq{}}\PY{p}{)}\PY{p}{)}

\PY{p}{(}\PY{n+nf}{defpolyfn} \PY{n+nv}{stmt2str} \PY{l+s+ss}{\PYZsq{}types.PrimitiveType} \PY{p}{[}\PY{o}{\PYZca{}}\PY{n+nv}{org.eclipse.emf.ecore.EObject} \PY{n+nv}{pt}\PY{p}{]}
  \PY{p}{(}\PY{n+nf}{clojure.string/lower\PYZhy{}case} \PY{p}{(}\PY{n+nf}{.getName} \PY{p}{(}\PY{n+nf}{.eClass} \PY{n+nv}{pt}\PY{p}{)}\PY{p}{)}\PY{p}{)}\PY{p}{)}

\PY{p}{(}\PY{n+nf}{defpolyfn} \PY{n+nv}{stmt2str} \PY{l+s+ss}{\PYZsq{}statements.LocalVariableStatement} \PY{p}{[}\PY{n+nv}{lv}\PY{p}{]}
  \PY{p}{(}\PY{k}{let }\PY{p}{[}\PY{n+nv}{v} \PY{p}{(}\PY{n+nf}{eget} \PY{n+nv}{lv} \PY{l+s+ss}{:variable}\PY{p}{)}\PY{p}{]}
    \PY{p}{(}\PY{n+nb}{str }\PY{p}{(}\PY{n+nf}{stmt2str} \PY{p}{(}\PY{n+nf}{eget} \PY{n+nv}{v} \PY{l+s+ss}{:typeReference}\PY{p}{)}\PY{p}{)} \PY{l+s}{\PYZdq{} \PYZdq{}} \PY{p}{(}\PY{n+nf}{stmt2str} \PY{n+nv}{v}\PY{p}{)}
         \PY{p}{(}\PY{n+nb}{when\PYZhy{}let }\PY{p}{[}\PY{n+nv}{iv} \PY{p}{(}\PY{n+nf}{eget} \PY{n+nv}{v} \PY{l+s+ss}{:initialValue}\PY{p}{)}\PY{p}{]}
           \PY{p}{(}\PY{n+nb}{str }\PY{l+s}{\PYZdq{} = \PYZdq{}} \PY{p}{(}\PY{n+nf}{stmt2str} \PY{n+nv}{iv}\PY{p}{)}\PY{p}{)}\PY{p}{)}
         \PY{l+s}{\PYZdq{};\PYZdq{}}\PY{p}{)}\PY{p}{)}\PY{p}{)}

\PY{p}{(}\PY{n+nf}{defpolyfn} \PY{n+nv}{stmt2str} \PY{l+s+ss}{\PYZsq{}references.IdentifierReference} \PY{p}{[}\PY{n+nv}{ir}\PY{p}{]}
  \PY{p}{(}\PY{n+nf}{stmt2str} \PY{p}{(}\PY{n+nf}{eget} \PY{n+nv}{ir} \PY{l+s+ss}{:target}\PY{p}{)}\PY{p}{)}\PY{p}{)}

\PY{p}{(}\PY{n+nf}{defpolyfn} \PY{n+nv}{stmt2str} \PY{l+s+ss}{\PYZsq{}variables.Variable} \PY{p}{[}\PY{n+nv}{v}\PY{p}{]}
  \PY{p}{(}\PY{n+nf}{eget} \PY{n+nv}{v} \PY{l+s+ss}{:name}\PY{p}{)}\PY{p}{)}

\PY{p}{(}\PY{n+nf}{defpolyfn} \PY{n+nv}{stmt2str} \PY{l+s+ss}{\PYZsq{}expressions.MultiplicativeExpression} \PY{p}{[}\PY{n+nv}{me}\PY{p}{]}
  \PY{p}{(}\PY{k}{let }\PY{p}{[}\PY{p}{[}\PY{n+nv}{c1} \PY{n+nv}{c2}\PY{p}{]} \PY{p}{(}\PY{n+nf}{eget} \PY{n+nv}{me} \PY{l+s+ss}{:children}\PY{p}{)}\PY{p}{]}
    \PY{p}{(}\PY{n+nb}{str }\PY{p}{(}\PY{n+nf}{stmt2str} \PY{n+nv}{c1}\PY{p}{)} \PY{l+s}{\PYZdq{} \PYZdq{}} \PY{p}{(}\PY{n+nf}{reduce\PYZhy{}str} \PY{p}{(}\PY{n+nf}{eget} \PY{n+nv}{me} \PY{l+s+ss}{:multiplicativeOperators}\PY{p}{)}\PY{p}{)}
         \PY{l+s}{\PYZdq{} \PYZdq{}} \PY{p}{(}\PY{n+nf}{stmt2str} \PY{n+nv}{c2}\PY{p}{)}\PY{p}{)}\PY{p}{)}\PY{p}{)}

\PY{p}{(}\PY{n+nf}{defpolyfn} \PY{n+nv}{stmt2str} \PY{l+s+ss}{\PYZsq{}expressions.EqualityExpression} \PY{p}{[}\PY{n+nv}{ee}\PY{p}{]}
  \PY{p}{(}\PY{k}{let }\PY{p}{[}\PY{p}{[}\PY{n+nv}{c1} \PY{n+nv}{c2}\PY{p}{]} \PY{p}{(}\PY{n+nf}{eget} \PY{n+nv}{ee} \PY{l+s+ss}{:children}\PY{p}{)}\PY{p}{]}
    \PY{p}{(}\PY{n+nb}{str }\PY{p}{(}\PY{n+nf}{stmt2str} \PY{n+nv}{c1}\PY{p}{)} \PY{l+s}{\PYZdq{} \PYZdq{}} \PY{p}{(}\PY{n+nf}{reduce\PYZhy{}str} \PY{p}{(}\PY{n+nf}{eget} \PY{n+nv}{ee} \PY{l+s+ss}{:equalityOperators}\PY{p}{)}\PY{p}{)}
         \PY{l+s}{\PYZdq{} \PYZdq{}} \PY{p}{(}\PY{n+nf}{stmt2str} \PY{n+nv}{c2}\PY{p}{)}\PY{p}{)}\PY{p}{)}\PY{p}{)}

\PY{p}{(}\PY{n+nf}{defpolyfn} \PY{n+nv}{stmt2str} \PY{l+s+ss}{\PYZsq{}expressions.AdditiveExpression} \PY{p}{[}\PY{n+nv}{ae}\PY{p}{]}
  \PY{p}{(}\PY{k}{let }\PY{p}{[}\PY{p}{[}\PY{n+nv}{c1} \PY{n+nv}{c2}\PY{p}{]} \PY{p}{(}\PY{n+nf}{eget} \PY{n+nv}{ae} \PY{l+s+ss}{:children}\PY{p}{)}\PY{p}{]}
    \PY{p}{(}\PY{n+nb}{str }\PY{p}{(}\PY{n+nf}{stmt2str} \PY{n+nv}{c1}\PY{p}{)} \PY{l+s}{\PYZdq{} \PYZdq{}} \PY{p}{(}\PY{n+nf}{reduce\PYZhy{}str} \PY{p}{(}\PY{n+nf}{eget} \PY{n+nv}{ae} \PY{l+s+ss}{:additiveOperators}\PY{p}{)}\PY{p}{)}
         \PY{l+s}{\PYZdq{} \PYZdq{}} \PY{p}{(}\PY{n+nf}{stmt2str} \PY{n+nv}{c2}\PY{p}{)}\PY{p}{)}\PY{p}{)}\PY{p}{)}

\PY{p}{(}\PY{n+nf}{defpolyfn} \PY{n+nv}{stmt2str} \PY{l+s+ss}{\PYZsq{}expressions.UnaryExpression} \PY{p}{[}\PY{n+nv}{ue}\PY{p}{]}
  \PY{p}{(}\PY{n+nb}{str }\PY{p}{(}\PY{n+nf}{reduce\PYZhy{}str} \PY{p}{(}\PY{n+nf}{eget} \PY{n+nv}{ue} \PY{l+s+ss}{:operators}\PY{p}{)}\PY{p}{)}
       \PY{p}{(}\PY{n+nf}{stmt2str} \PY{p}{(}\PY{n+nf}{eget} \PY{n+nv}{ue} \PY{l+s+ss}{:child}\PY{p}{)}\PY{p}{)}\PY{p}{)}\PY{p}{)}

\PY{p}{(}\PY{n+nf}{defpolyfn} \PY{n+nv}{stmt2str} \PY{l+s+ss}{\PYZsq{}expressions.AssignmentExpression} \PY{p}{[}\PY{n+nv}{ae}\PY{p}{]}
  \PY{p}{(}\PY{n+nb}{str }\PY{p}{(}\PY{n+nf}{stmt2str} \PY{p}{(}\PY{n+nf}{eget} \PY{n+nv}{ae} \PY{l+s+ss}{:child}\PY{p}{)}\PY{p}{)} \PY{l+s}{\PYZdq{} \PYZdq{}}
       \PY{p}{(}\PY{n+nf}{stmt2str} \PY{p}{(}\PY{n+nf}{eget} \PY{n+nv}{ae} \PY{l+s+ss}{:assignmentOperator}\PY{p}{)}\PY{p}{)} \PY{l+s}{\PYZdq{} \PYZdq{}}
       \PY{p}{(}\PY{n+nf}{stmt2str} \PY{p}{(}\PY{n+nf}{eget} \PY{n+nv}{ae} \PY{l+s+ss}{:value}\PY{p}{)}\PY{p}{)}\PY{p}{)}\PY{p}{)}

\PY{p}{(}\PY{n+nf}{defpolyfn} \PY{n+nv}{stmt2str} \PY{l+s+ss}{\PYZsq{}expressions.RelationExpression} \PY{p}{[}\PY{n+nv}{re}\PY{p}{]}
  \PY{p}{(}\PY{k}{let }\PY{p}{[}\PY{p}{[}\PY{n+nv}{c1} \PY{n+nv}{c2}\PY{p}{]} \PY{p}{(}\PY{n+nf}{eget} \PY{n+nv}{re} \PY{l+s+ss}{:children}\PY{p}{)}\PY{p}{]}
    \PY{p}{(}\PY{n+nb}{str }\PY{p}{(}\PY{n+nf}{stmt2str} \PY{n+nv}{c1}\PY{p}{)} \PY{l+s}{\PYZdq{} \PYZdq{}}
         \PY{p}{(}\PY{n+nf}{reduce\PYZhy{}str} \PY{p}{(}\PY{n+nf}{eget} \PY{n+nv}{re} \PY{l+s+ss}{:relationOperators}\PY{p}{)}\PY{p}{)}
         \PY{l+s}{\PYZdq{} \PYZdq{}} \PY{p}{(}\PY{n+nf}{stmt2str} \PY{n+nv}{c2}\PY{p}{)}\PY{p}{)}\PY{p}{)}\PY{p}{)}

\PY{p}{(}\PY{n+nf}{defpolyfn} \PY{n+nv}{stmt2str} \PY{l+s+ss}{\PYZsq{}expressions.SuffixUnaryModificationExpression} \PY{p}{[}\PY{n+nv}{se}\PY{p}{]}
  \PY{p}{(}\PY{n+nb}{str }\PY{p}{(}\PY{n+nf}{stmt2str} \PY{p}{(}\PY{n+nf}{eget} \PY{n+nv}{se} \PY{l+s+ss}{:child}\PY{p}{)}\PY{p}{)}
       \PY{p}{(}\PY{n+nf}{stmt2str} \PY{p}{(}\PY{n+nf}{eget} \PY{n+nv}{se} \PY{l+s+ss}{:operator}\PY{p}{)}\PY{p}{)}\PY{p}{)}\PY{p}{)}

\PY{p}{(}\PY{n+nf}{defpolyfn} \PY{n+nv}{stmt2str} \PY{l+s+ss}{\PYZsq{}statements.Block} \PY{p}{[}\PY{n+nv}{b}\PY{p}{]}
  \PY{l+s}{\PYZdq{}\PYZbs{}\PYZbs{}\PYZob{}...\PYZbs{}\PYZbs{}\PYZcb{}\PYZdq{}}\PY{p}{)}

\PY{p}{(}\PY{n+nf}{defpolyfn} \PY{n+nv}{stmt2str} \PY{l+s+ss}{\PYZsq{}statements.Condition} \PY{p}{[}\PY{n+nv}{c}\PY{p}{]}
  \PY{l+s}{\PYZdq{}if\PYZdq{}}\PY{p}{)}

\PY{p}{(}\PY{n+nf}{defpolyfn} \PY{n+nv}{stmt2str} \PY{l+s+ss}{\PYZsq{}statements.WhileLoop} \PY{p}{[}\PY{n+nv}{c}\PY{p}{]}
  \PY{l+s}{\PYZdq{}while\PYZdq{}}\PY{p}{)}

\PY{p}{(}\PY{n+nf}{defpolyfn} \PY{n+nv}{stmt2str} \PY{l+s+ss}{\PYZsq{}statements.JumpLabel} \PY{p}{[}\PY{n+nv}{l}\PY{p}{]}
  \PY{p}{(}\PY{n+nb}{str }\PY{p}{(}\PY{n+nf}{eget} \PY{n+nv}{l} \PY{l+s+ss}{:name}\PY{p}{)} \PY{l+s}{\PYZdq{}:\PYZdq{}}\PY{p}{)}\PY{p}{)}

\PY{p}{(}\PY{n+nf}{defpolyfn} \PY{n+nv}{stmt2str} \PY{l+s+ss}{\PYZsq{}statements.Break} \PY{p}{[}\PY{n+nv}{b}\PY{p}{]}
  \PY{p}{(}\PY{n+nb}{str }\PY{l+s}{\PYZdq{}break\PYZdq{}}
       \PY{p}{(}\PY{n+nb}{when\PYZhy{}let }\PY{p}{[}\PY{n+nv}{l} \PY{p}{(}\PY{n+nf}{eget} \PY{n+nv}{b} \PY{l+s+ss}{:target}\PY{p}{)}\PY{p}{]}
         \PY{p}{(}\PY{n+nb}{str }\PY{l+s}{\PYZdq{} \PYZdq{}} \PY{p}{(}\PY{n+nf}{eget} \PY{n+nv}{l} \PY{l+s+ss}{:name}\PY{p}{)}\PY{p}{)}\PY{p}{)}
       \PY{l+s}{\PYZdq{};\PYZdq{}}\PY{p}{)}\PY{p}{)}

\PY{p}{(}\PY{n+nf}{defpolyfn} \PY{n+nv}{stmt2str} \PY{l+s+ss}{\PYZsq{}statements.Continue} \PY{p}{[}\PY{n+nv}{c}\PY{p}{]}
  \PY{p}{(}\PY{n+nb}{str }\PY{l+s}{\PYZdq{}continue\PYZdq{}}
       \PY{p}{(}\PY{n+nb}{when\PYZhy{}let }\PY{p}{[}\PY{n+nv}{l} \PY{p}{(}\PY{n+nf}{eget} \PY{n+nv}{c} \PY{l+s+ss}{:target}\PY{p}{)}\PY{p}{]}
         \PY{p}{(}\PY{n+nb}{str }\PY{l+s}{\PYZdq{} \PYZdq{}} \PY{p}{(}\PY{n+nf}{eget} \PY{n+nv}{l} \PY{l+s+ss}{:name}\PY{p}{)}\PY{p}{)}\PY{p}{)}
       \PY{l+s}{\PYZdq{};\PYZdq{}}\PY{p}{)}\PY{p}{)}

\PY{p}{(}\PY{n+nf}{defpolyfn} \PY{n+nv}{stmt2str} \PY{l+s+ss}{\PYZsq{}statements.Return} \PY{p}{[}\PY{n+nv}{r}\PY{p}{]}
  \PY{p}{(}\PY{n+nb}{str }\PY{l+s}{\PYZdq{}return\PYZdq{}} \PY{p}{(}\PY{n+nb}{when\PYZhy{}let }\PY{p}{[}\PY{n+nv}{rv} \PY{p}{(}\PY{n+nf}{eget} \PY{n+nv}{r} \PY{l+s+ss}{:returnValue}\PY{p}{)}\PY{p}{]}
                  \PY{p}{(}\PY{n+nb}{str }\PY{l+s}{\PYZdq{} \PYZdq{}} \PY{p}{(}\PY{n+nf}{stmt2str} \PY{n+nv}{rv}\PY{p}{)}\PY{p}{)}\PY{p}{)}
       \PY{l+s}{\PYZdq{};\PYZdq{}}\PY{p}{)}\PY{p}{)}

\PY{p}{(}\PY{n+nf}{defpolyfn} \PY{n+nv}{stmt2str} \PY{l+s+ss}{\PYZsq{}statements.ExpressionStatement} \PY{p}{[}\PY{n+nv}{stmt}\PY{p}{]}
  \PY{p}{(}\PY{n+nb}{str }\PY{p}{(}\PY{n+nf}{stmt2str} \PY{p}{(}\PY{n+nf}{eget} \PY{n+nv}{stmt} \PY{l+s+ss}{:expression}\PY{p}{)}\PY{p}{)} \PY{l+s}{\PYZdq{};\PYZdq{}}\PY{p}{)}\PY{p}{)}

\PY{p}{(}\PY{n+nf}{defpolyfn} \PY{n+nv}{stmt2str} \PY{l+s+ss}{\PYZsq{}operators.Operator} \PY{p}{[}\PY{n+nv}{op}\PY{p}{]}
  \PY{p}{(}\PY{n+nf}{type\PYZhy{}case} \PY{n+nv}{op}
    \PY{l+s+ss}{\PYZsq{}operators.Multiplication} \PY{l+s}{\PYZdq{}*\PYZdq{}}
    \PY{l+s+ss}{\PYZsq{}operators.Subtraction}    \PY{l+s}{\PYZdq{}\PYZhy{}\PYZdq{}}
    \PY{l+s+ss}{\PYZsq{}operators.Addition}       \PY{l+s}{\PYZdq{}+\PYZdq{}}
    \PY{l+s+ss}{\PYZsq{}operators.Division}       \PY{l+s}{\PYZdq{}/\PYZdq{}}
    \PY{l+s+ss}{\PYZsq{}operators.LessThan}       \PY{l+s}{\PYZdq{}\PYZlt{}\PYZdq{}}
    \PY{l+s+ss}{\PYZsq{}operators.GreaterThan}    \PY{l+s}{\PYZdq{}\PYZgt{}\PYZdq{}}
    \PY{l+s+ss}{\PYZsq{}operators.Assignment}     \PY{l+s}{\PYZdq{}=\PYZdq{}}
    \PY{l+s+ss}{\PYZsq{}operators.MinusMinus}     \PY{l+s}{\PYZdq{}\PYZhy{}\PYZhy{}\PYZdq{}}
    \PY{l+s+ss}{\PYZsq{}operators.PlusPlus}       \PY{l+s}{\PYZdq{}++\PYZdq{}}
    \PY{l+s+ss}{\PYZsq{}operators.AssignmentPlus} \PY{l+s}{\PYZdq{}+=\PYZdq{}}
    \PY{l+s+ss}{\PYZsq{}operators.Equal}          \PY{l+s}{\PYZdq{}==\PYZdq{}}\PY{p}{)}\PY{p}{)}

\PY{p}{(}\PY{n+nf}{defpolyfn} \PY{n+nv}{stmt2str} \PY{l+s+ss}{\PYZsq{}literals.Literal} \PY{p}{[}\PY{n+nv}{l}\PY{p}{]}
  \PY{p}{(}\PY{n+nf}{type\PYZhy{}case} \PY{n+nv}{l}
    \PY{l+s+ss}{\PYZsq{}literals.DecimalIntegerLiteral} \PY{p}{(}\PY{n+nf}{eget} \PY{n+nv}{l} \PY{l+s+ss}{:decimalValue}\PY{p}{)}\PY{p}{)}\PY{p}{)}
\end{Verbatim}

\section{The complete JaMoPP-to-StructureGraph Transformation}
\label{sec:compl-jamopp-struct}

\begin{Verbatim}[commandchars=\\\{\},fontsize=\footnotesize]
\PY{p}{(}\PY{k+kd}{defn }\PY{n+nv}{used\PYZhy{}vars} \PY{p}{[}\PY{n+nv}{s}\PY{p}{]}
  \PY{p}{(}\PY{n+nf}{reachables} \PY{n+nv}{s} \PY{p}{[}\PY{n+nv}{p\PYZhy{}seq} \PY{p}{[}\PY{n+nv}{p\PYZhy{}*} \PY{n+nv}{\PYZlt{}\PYZgt{}\PYZhy{}\PYZhy{}}\PY{p}{]}
                 \PY{p}{[}\PY{n+nv}{p\PYZhy{}restr} \PY{l+s+ss}{\PYZsq{}references.IdentifierReference}\PY{p}{]}
                 \PY{l+s+ss}{:target}\PY{p}{]}\PY{p}{)}\PY{p}{)}

\PY{p}{(}\PY{n+nf}{deftransformation} \PY{n+nv}{java2flowgraph} \PY{p}{[}\PY{p}{[}\PY{n+nv}{in} \PY{l+s+ss}{:emf}\PY{p}{]} \PY{p}{[}\PY{n+nv}{out} \PY{l+s+ss}{:emf}\PY{p}{]}\PY{p}{]}
  \PY{p}{(}\PY{o}{\PYZca{}}\PY{l+s+ss}{:top} \PY{n+nv}{method2method} \PY{p}{[}\PY{n+nv}{m}\PY{p}{]}
      \PY{l+s+ss}{:from} \PY{l+s+ss}{\PYZsq{}members.ClassMethod}
      \PY{l+s+ss}{:to} \PY{p}{[}\PY{n+nv}{fgm} \PY{l+s+ss}{\PYZsq{}flowgraph.Method}, \PY{n+nv}{fgex} \PY{l+s+ss}{\PYZsq{}flowgraph.Exit}\PY{p}{]}
      \PY{p}{(}\PY{n+nf}{eset!} \PY{n+nv}{fgm} \PY{l+s+ss}{:txt} \PY{p}{(}\PY{n+nf}{stmt2str} \PY{n+nv}{m}\PY{p}{)}\PY{p}{)}
      \PY{p}{(}\PY{n+nf}{eset!} \PY{n+nv}{fgex} \PY{l+s+ss}{:txt} \PY{l+s}{\PYZdq{}Exit\PYZdq{}}\PY{p}{)}
      \PY{p}{(}\PY{n+nf}{eset!} \PY{n+nv}{fgm} \PY{l+s+ss}{:stmts} \PY{p}{(}\PY{n+nb}{map }\PY{n+nv}{stmt2item} \PY{p}{(}\PY{n+nb}{seq }\PY{p}{(}\PY{n+nf}{eget} \PY{n+nv}{m} \PY{l+s+ss}{:statements}\PY{p}{)}\PY{p}{)}\PY{p}{)}\PY{p}{)}
      \PY{p}{(}\PY{n+nf}{eset!} \PY{n+nv}{fgm} \PY{l+s+ss}{:exit} \PY{n+nv}{fgex}\PY{p}{)}
      \PY{p}{(}\PY{n+nf}{eset!} \PY{n+nv}{fgm} \PY{l+s+ss}{:def} \PY{p}{(}\PY{n+nb}{map }\PY{n+nv}{param2param} \PY{p}{(}\PY{n+nf}{eget} \PY{n+nv}{m} \PY{l+s+ss}{:parameters}\PY{p}{)}\PY{p}{)}\PY{p}{)}\PY{p}{)}
  \PY{p}{(}\PY{n+nf}{stmt2item} \PY{p}{[}\PY{n+nv}{stmt}\PY{p}{]}
      \PY{l+s+ss}{:generalizes} \PY{p}{[}\PY{n+nv}{local\PYZhy{}var\PYZhy{}stmt2simple\PYZhy{}stmt} \PY{n+nv}{condition2if} \PY{n+nv}{block2block}
                    \PY{n+nv}{return2return} \PY{n+nv}{while\PYZhy{}loop2loop} \PY{n+nv}{break2break} \PY{n+nv}{continue2continue}
                    \PY{n+nv}{label2label} \PY{n+nv}{stmt2simple\PYZhy{}stmt}\PY{p}{]}\PY{p}{)}
  \PY{p}{(}\PY{n+nf}{var\PYZhy{}creating\PYZhy{}rule} \PY{p}{[}\PY{n+nv}{v}\PY{p}{]}
      \PY{l+s+ss}{:generalizes} \PY{p}{[}\PY{n+nv}{param2param} \PY{n+nv}{local\PYZhy{}var2var}\PY{p}{]}\PY{p}{)}
  \PY{p}{(}\PY{n+nf}{param2param} \PY{p}{[}\PY{n+nv}{p}\PY{p}{]}
      \PY{l+s+ss}{:from} \PY{l+s+ss}{\PYZsq{}parameters.Parameter}
      \PY{l+s+ss}{:to} \PY{p}{[}\PY{n+nv}{fgp} \PY{l+s+ss}{\PYZsq{}flowgraph.Param}\PY{p}{]}
      \PY{p}{(}\PY{n+nf}{eset!} \PY{n+nv}{fgp} \PY{l+s+ss}{:txt} \PY{p}{(}\PY{n+nf}{stmt2str} \PY{n+nv}{p}\PY{p}{)}\PY{p}{)}\PY{p}{)}
  \PY{p}{(}\PY{n+nf}{local\PYZhy{}var2var} \PY{p}{[}\PY{n+nv}{lv}\PY{p}{]}
      \PY{l+s+ss}{:from} \PY{l+s+ss}{\PYZsq{}variables.LocalVariable}
      \PY{l+s+ss}{:to} \PY{p}{[}\PY{n+nv}{fgv} \PY{l+s+ss}{\PYZsq{}flowgraph.Var}\PY{p}{]}
      \PY{p}{(}\PY{n+nf}{eset!} \PY{n+nv}{fgv} \PY{l+s+ss}{:txt} \PY{p}{(}\PY{n+nf}{stmt2str} \PY{n+nv}{lv}\PY{p}{)}\PY{p}{)}\PY{p}{)}
  \PY{p}{(}\PY{n+nf}{local\PYZhy{}var\PYZhy{}stmt2simple\PYZhy{}stmt} \PY{p}{[}\PY{n+nv}{lv}\PY{p}{]}
      \PY{l+s+ss}{:from} \PY{l+s+ss}{\PYZsq{}statements.LocalVariableStatement}
      \PY{l+s+ss}{:to} \PY{p}{[}\PY{n+nv}{fgss} \PY{l+s+ss}{\PYZsq{}flowgraph.SimpleStmt}\PY{p}{]}
      \PY{p}{(}\PY{k}{let }\PY{p}{[}\PY{n+nv}{v} \PY{p}{(}\PY{n+nf}{local\PYZhy{}var2var} \PY{p}{(}\PY{n+nf}{adj} \PY{n+nv}{lv} \PY{l+s+ss}{:variable}\PY{p}{)}\PY{p}{)}\PY{p}{]}
        \PY{p}{(}\PY{n+nf}{eset!} \PY{n+nv}{fgss} \PY{l+s+ss}{:txt} \PY{p}{(}\PY{n+nf}{stmt2str} \PY{n+nv}{lv}\PY{p}{)}\PY{p}{)}
        \PY{p}{(}\PY{n+nf}{eadd!} \PY{n+nv}{fgss} \PY{l+s+ss}{:def} \PY{n+nv}{v}\PY{p}{)}
        \PY{p}{(}\PY{n+nf}{eset!} \PY{n+nv}{fgss} \PY{l+s+ss}{:use} \PY{p}{(}\PY{n+nb}{map }\PY{n+nv}{var\PYZhy{}creating\PYZhy{}rule}
                              \PY{p}{(}\PY{n+nf}{used\PYZhy{}vars} \PY{p}{(}\PY{n+nf}{adj} \PY{n+nv}{lv} \PY{l+s+ss}{:variable} \PY{l+s+ss}{:initialValue}\PY{p}{)}\PY{p}{)}\PY{p}{)}\PY{p}{)}\PY{p}{)}\PY{p}{)}
  \PY{p}{(}\PY{n+nf}{stmt2simple\PYZhy{}stmt} \PY{p}{[}\PY{n+nv}{s}\PY{p}{]}
      \PY{l+s+ss}{:from} \PY{l+s+ss}{\PYZsq{}statements.Statement}
      \PY{l+s+ss}{:to} \PY{p}{[}\PY{n+nv}{fgss} \PY{l+s+ss}{\PYZsq{}flowgraph.SimpleStmt}\PY{p}{]}
      \PY{p}{(}\PY{n+nf}{eset!} \PY{n+nv}{fgss} \PY{l+s+ss}{:txt} \PY{p}{(}\PY{n+nf}{stmt2str} \PY{n+nv}{s}\PY{p}{)}\PY{p}{)}
      \PY{p}{(}\PY{n+nb}{doseq }\PY{p}{[}\PY{n+nv}{aex}  \PY{p}{(}\PY{n+nf}{reachables} \PY{n+nv}{s} \PY{p}{[}\PY{n+nv}{p\PYZhy{}seq} \PY{p}{[}\PY{n+nv}{p\PYZhy{}*} \PY{n+nv}{\PYZlt{}\PYZgt{}\PYZhy{}\PYZhy{}}\PY{p}{]}
                                  \PY{p}{[}\PY{n+nv}{p\PYZhy{}restr} \PY{l+s+ss}{\PYZsq{}expressions.AssignmentExpression}\PY{p}{]}\PY{p}{]}\PY{p}{)}\PY{p}{]}
        \PY{p}{(}\PY{n+nf}{eadd!} \PY{n+nv}{fgss} \PY{l+s+ss}{:def} \PY{p}{(}\PY{n+nf}{var\PYZhy{}creating\PYZhy{}rule} \PY{p}{(}\PY{n+nf}{the} \PY{p}{(}\PY{n+nf}{used\PYZhy{}vars} \PY{p}{(}\PY{n+nf}{adj} \PY{n+nv}{aex} \PY{l+s+ss}{:child}\PY{p}{)}\PY{p}{)}\PY{p}{)}\PY{p}{)}\PY{p}{)}
        \PY{p}{(}\PY{n+nf}{eaddall!} \PY{n+nv}{fgss} \PY{l+s+ss}{:use} \PY{p}{(}\PY{n+nb}{map }\PY{n+nv}{var\PYZhy{}creating\PYZhy{}rule} \PY{p}{(}\PY{n+nf}{used\PYZhy{}vars} \PY{p}{(}\PY{n+nf}{adj} \PY{n+nv}{aex} \PY{l+s+ss}{:value}\PY{p}{)}\PY{p}{)}\PY{p}{)}\PY{p}{)}\PY{p}{)}
      \PY{p}{(}\PY{n+nb}{doseq }\PY{p}{[}\PY{n+nv}{umex} \PY{p}{(}\PY{n+nf}{reachables} \PY{n+nv}{s} \PY{p}{[}\PY{n+nv}{p\PYZhy{}seq} \PY{p}{[}\PY{n+nv}{p\PYZhy{}*} \PY{n+nv}{\PYZlt{}\PYZgt{}\PYZhy{}\PYZhy{}}\PY{p}{]}
                                  \PY{p}{[}\PY{n+nv}{p\PYZhy{}restr} \PY{l+s+ss}{\PYZsq{}expressions.UnaryModificationExpression}\PY{p}{]}\PY{p}{]}\PY{p}{)}\PY{p}{]}
        \PY{p}{(}\PY{k}{let }\PY{p}{[}\PY{k}{var }\PY{p}{(}\PY{n+nf}{var\PYZhy{}creating\PYZhy{}rule} \PY{p}{(}\PY{n+nf}{the} \PY{p}{(}\PY{n+nf}{used\PYZhy{}vars} \PY{p}{(}\PY{n+nf}{adj} \PY{n+nv}{umex} \PY{l+s+ss}{:child}\PY{p}{)}\PY{p}{)}\PY{p}{)}\PY{p}{)}\PY{p}{]}
          \PY{p}{(}\PY{n+nf}{eadd!} \PY{n+nv}{fgss} \PY{l+s+ss}{:def} \PY{n+nv}{var}\PY{p}{)}
          \PY{p}{(}\PY{n+nf}{eadd!} \PY{n+nv}{fgss} \PY{l+s+ss}{:use} \PY{n+nv}{var}\PY{p}{)}\PY{p}{)}\PY{p}{)}\PY{p}{)}
  \PY{p}{(}\PY{n+nf}{label2label} \PY{p}{[}\PY{n+nv}{l}\PY{p}{]}
      \PY{l+s+ss}{:from} \PY{l+s+ss}{\PYZsq{}statements.JumpLabel}
      \PY{l+s+ss}{:to} \PY{p}{[}\PY{n+nv}{fgl} \PY{l+s+ss}{\PYZsq{}flowgraph.Label}\PY{p}{]}
      \PY{p}{(}\PY{n+nf}{eset!} \PY{n+nv}{fgl} \PY{l+s+ss}{:txt} \PY{p}{(}\PY{n+nf}{stmt2str} \PY{n+nv}{l}\PY{p}{)}\PY{p}{)}
      \PY{p}{(}\PY{n+nf}{eset!} \PY{n+nv}{fgl} \PY{l+s+ss}{:stmt} \PY{p}{(}\PY{n+nf}{stmt2item} \PY{p}{(}\PY{n+nf}{eget} \PY{n+nv}{l} \PY{l+s+ss}{:statement}\PY{p}{)}\PY{p}{)}\PY{p}{)}\PY{p}{)}
  \PY{p}{(}\PY{n+nf}{expression2expr} \PY{p}{[}\PY{n+nv}{ex}\PY{p}{]}
      \PY{l+s+ss}{:from} \PY{l+s+ss}{\PYZsq{}expressions.Expression}
      \PY{l+s+ss}{:to} \PY{p}{[}\PY{n+nv}{fgex} \PY{l+s+ss}{\PYZsq{}flowgraph.Expr}\PY{p}{]}
      \PY{p}{(}\PY{n+nf}{eset!} \PY{n+nv}{fgex} \PY{l+s+ss}{:txt} \PY{p}{(}\PY{n+nf}{stmt2str} \PY{n+nv}{ex}\PY{p}{)}\PY{p}{)}
      \PY{p}{(}\PY{n+nf}{eset!} \PY{n+nv}{fgex} \PY{l+s+ss}{:use} \PY{p}{(}\PY{n+nb}{map }\PY{n+nv}{var\PYZhy{}creating\PYZhy{}rule} \PY{p}{(}\PY{n+nf}{used\PYZhy{}vars} \PY{n+nv}{ex}\PY{p}{)}\PY{p}{)}\PY{p}{)}\PY{p}{)}
  \PY{p}{(}\PY{n+nf}{condition2if} \PY{p}{[}\PY{n+nv}{c}\PY{p}{]}
      \PY{l+s+ss}{:from} \PY{l+s+ss}{\PYZsq{}statements.Condition}
      \PY{l+s+ss}{:to} \PY{p}{[}\PY{n+nv}{fgif} \PY{l+s+ss}{\PYZsq{}flowgraph.If}\PY{p}{]}
      \PY{p}{(}\PY{n+nf}{eset!} \PY{n+nv}{fgif} \PY{l+s+ss}{:txt} \PY{p}{(}\PY{n+nf}{stmt2str} \PY{n+nv}{c}\PY{p}{)}\PY{p}{)}
      \PY{p}{(}\PY{n+nf}{eset!} \PY{n+nv}{fgif} \PY{l+s+ss}{:expr} \PY{p}{(}\PY{n+nf}{expression2expr} \PY{p}{(}\PY{n+nf}{eget} \PY{n+nv}{c} \PY{l+s+ss}{:condition}\PY{p}{)}\PY{p}{)}\PY{p}{)}
      \PY{p}{(}\PY{n+nf}{eset!} \PY{n+nv}{fgif} \PY{l+s+ss}{:then} \PY{p}{(}\PY{n+nf}{stmt2item} \PY{p}{(}\PY{n+nf}{eget} \PY{n+nv}{c} \PY{l+s+ss}{:statement}\PY{p}{)}\PY{p}{)}\PY{p}{)}
      \PY{p}{(}\PY{n+nb}{when\PYZhy{}let }\PY{p}{[}\PY{n+nv}{else} \PY{p}{(}\PY{n+nf}{eget} \PY{n+nv}{c} \PY{l+s+ss}{:elseStatement}\PY{p}{)}\PY{p}{]}
        \PY{p}{(}\PY{n+nf}{eset!} \PY{n+nv}{fgif} \PY{l+s+ss}{:else} \PY{p}{(}\PY{n+nf}{stmt2item} \PY{n+nv}{else}\PY{p}{)}\PY{p}{)}\PY{p}{)}\PY{p}{)}
  \PY{p}{(}\PY{n+nf}{block2block} \PY{p}{[}\PY{n+nv}{b}\PY{p}{]}
      \PY{l+s+ss}{:from} \PY{l+s+ss}{\PYZsq{}statements.Block}
      \PY{l+s+ss}{:to} \PY{p}{[}\PY{n+nv}{fgb} \PY{l+s+ss}{\PYZsq{}flowgraph.Block}\PY{p}{]}
      \PY{p}{(}\PY{n+nf}{eset!} \PY{n+nv}{fgb} \PY{l+s+ss}{:txt} \PY{p}{(}\PY{n+nf}{stmt2str} \PY{n+nv}{b}\PY{p}{)}\PY{p}{)}
      \PY{p}{(}\PY{n+nf}{eset!} \PY{n+nv}{fgb} \PY{l+s+ss}{:stmts} \PY{p}{(}\PY{n+nb}{map }\PY{n+nv}{stmt2item} \PY{p}{(}\PY{n+nf}{eget} \PY{n+nv}{b} \PY{l+s+ss}{:statements}\PY{p}{)}\PY{p}{)}\PY{p}{)}\PY{p}{)}
  \PY{p}{(}\PY{n+nf}{return2return} \PY{p}{[}\PY{n+nv}{r}\PY{p}{]}
      \PY{l+s+ss}{:from} \PY{l+s+ss}{\PYZsq{}statements.Return}
      \PY{l+s+ss}{:to} \PY{p}{[}\PY{n+nv}{fgr} \PY{l+s+ss}{\PYZsq{}flowgraph.Return}\PY{p}{]}
      \PY{p}{(}\PY{n+nf}{eset!} \PY{n+nv}{fgr} \PY{l+s+ss}{:txt} \PY{p}{(}\PY{n+nf}{stmt2str} \PY{n+nv}{r}\PY{p}{)}\PY{p}{)}
      \PY{p}{(}\PY{n+nf}{eset!} \PY{n+nv}{fgr} \PY{l+s+ss}{:use} \PY{p}{(}\PY{n+nb}{map }\PY{n+nv}{var\PYZhy{}creating\PYZhy{}rule} \PY{p}{(}\PY{n+nf}{used\PYZhy{}vars} \PY{n+nv}{r}\PY{p}{)}\PY{p}{)}\PY{p}{)}\PY{p}{)}
  \PY{p}{(}\PY{n+nf}{break2break} \PY{p}{[}\PY{n+nv}{b}\PY{p}{]}
      \PY{l+s+ss}{:from} \PY{l+s+ss}{\PYZsq{}statements.Break}
      \PY{l+s+ss}{:to} \PY{p}{[}\PY{n+nv}{fgb} \PY{l+s+ss}{\PYZsq{}flowgraph.Break}\PY{p}{]}
      \PY{p}{(}\PY{n+nf}{eset!} \PY{n+nv}{fgb} \PY{l+s+ss}{:txt} \PY{p}{(}\PY{n+nf}{stmt2str} \PY{n+nv}{b}\PY{p}{)}\PY{p}{)}
      \PY{p}{(}\PY{n+nf}{eset!} \PY{n+nv}{fgb} \PY{l+s+ss}{:label} \PY{p}{(}\PY{n+nf}{label2label} \PY{p}{(}\PY{n+nf}{eget} \PY{n+nv}{b} \PY{l+s+ss}{:target}\PY{p}{)}\PY{p}{)}\PY{p}{)}\PY{p}{)}
  \PY{p}{(}\PY{n+nf}{continue2continue} \PY{p}{[}\PY{n+nv}{c}\PY{p}{]}
      \PY{l+s+ss}{:from} \PY{l+s+ss}{\PYZsq{}statements.Continue}
      \PY{l+s+ss}{:to} \PY{p}{[}\PY{n+nv}{fgc} \PY{l+s+ss}{\PYZsq{}flowgraph.Continue}\PY{p}{]}
      \PY{p}{(}\PY{n+nf}{eset!} \PY{n+nv}{fgc} \PY{l+s+ss}{:txt} \PY{p}{(}\PY{n+nf}{stmt2str} \PY{n+nv}{c}\PY{p}{)}\PY{p}{)}
      \PY{p}{(}\PY{n+nf}{eset!} \PY{n+nv}{fgc} \PY{l+s+ss}{:label} \PY{p}{(}\PY{n+nf}{label2label} \PY{p}{(}\PY{n+nf}{eget} \PY{n+nv}{c} \PY{l+s+ss}{:target}\PY{p}{)}\PY{p}{)}\PY{p}{)}\PY{p}{)}
  \PY{p}{(}\PY{n+nf}{while\PYZhy{}loop2loop} \PY{p}{[}\PY{n+nv}{wl}\PY{p}{]}
      \PY{l+s+ss}{:from} \PY{l+s+ss}{\PYZsq{}statements.WhileLoop}
      \PY{l+s+ss}{:to}   \PY{p}{[}\PY{n+nv}{fgl} \PY{l+s+ss}{\PYZsq{}flowgraph.Loop}\PY{p}{]}
      \PY{p}{(}\PY{n+nf}{eset!} \PY{n+nv}{fgl} \PY{l+s+ss}{:txt} \PY{p}{(}\PY{n+nf}{stmt2str} \PY{n+nv}{wl}\PY{p}{)}\PY{p}{)}
      \PY{p}{(}\PY{n+nf}{eset!} \PY{n+nv}{fgl} \PY{l+s+ss}{:expr} \PY{p}{(}\PY{n+nf}{expression2expr} \PY{p}{(}\PY{n+nf}{eget} \PY{n+nv}{wl} \PY{l+s+ss}{:condition}\PY{p}{)}\PY{p}{)}\PY{p}{)}
      \PY{p}{(}\PY{n+nf}{eset!} \PY{n+nv}{fgl} \PY{l+s+ss}{:body} \PY{p}{(}\PY{n+nf}{stmt2item} \PY{p}{(}\PY{n+nf}{eget} \PY{n+nv}{wl} \PY{l+s+ss}{:statement}\PY{p}{)}\PY{p}{)}\PY{p}{)}\PY{p}{)}\PY{p}{)}
\end{Verbatim}

\section{The complete Control Flow Transformation}
\label{sec:compl-contr-flow}

\begin{Verbatim}[commandchars=\\\{\},fontsize=\footnotesize]
\PY{p}{(}\PY{k+kd}{defn }\PY{n+nv}{cf\PYZhy{}peek} \PY{p}{[}\PY{n+nv}{el}\PY{p}{]}
  \PY{p}{(}\PY{k}{if }\PY{p}{(}\PY{n+nf}{has\PYZhy{}type?} \PY{n+nv}{el} \PY{l+s+ss}{\PYZsq{}flowgraph.FlowInstr}\PY{p}{)}
    \PY{n+nv}{el}
    \PY{p}{(}\PY{n+nf}{recur} \PY{p}{(}\PY{n+nb}{first }\PY{p}{(}\PY{n+nf}{econtents} \PY{n+nv}{el}\PY{p}{)}\PY{p}{)}\PY{p}{)}\PY{p}{)}\PY{p}{)}

\PY{p}{(}\PY{k+kd}{defn }\PY{n+nv}{cf\PYZhy{}synth} \PY{p}{[}\PY{n+nv}{v} \PY{n+nv}{exit} \PY{n+nv}{loop\PYZhy{}expr} \PY{n+nv}{loop\PYZhy{}succ} \PY{n+nv}{label\PYZhy{}succ\PYZhy{}map}\PY{p}{]}
  \PY{p}{(}\PY{n+nb}{when }\PY{p}{(}\PY{n+nb}{seq }\PY{n+nv}{v}\PY{p}{)}
    \PY{p}{(}\PY{k}{let }\PY{p}{[}\PY{p}{[}\PY{n+nv}{el} \PY{o}{\PYZam{}} \PY{p}{[}\PY{n+nv}{n} \PY{o}{\PYZam{}} \PY{n+nv}{\PYZus{}} \PY{l+s+ss}{:as} \PY{n+nv}{tail}\PY{p}{]}\PY{p}{]} \PY{n+nv}{v}\PY{p}{]}
      \PY{p}{(}\PY{n+nf}{type\PYZhy{}case} \PY{n+nv}{el}
        \PY{l+s+ss}{\PYZsq{}flowgraph.Method}
                   \PY{p}{(}\PY{k}{let }\PY{p}{[}\PY{n+nv}{stmts} \PY{p}{(}\PY{n+nf}{econtents} \PY{n+nv}{el}\PY{p}{)}\PY{p}{]}
                     \PY{p}{(}\PY{n+nf}{eadd!} \PY{n+nv}{el} \PY{l+s+ss}{:cfNext} \PY{p}{(}\PY{n+nf}{cf\PYZhy{}peek} \PY{p}{(}\PY{n+nb}{first }\PY{n+nv}{stmts}\PY{p}{)}\PY{p}{)}\PY{p}{)}
                     \PY{p}{(}\PY{n+nf}{recur} \PY{n+nv}{stmts} \PY{n+nv}{exit} \PY{n+nv}{nil} \PY{n+nv}{nil} \PY{n+nv}{nil}\PY{p}{)}\PY{p}{)}
        \PY{l+s+ss}{\PYZsq{}flowgraph.SimpleStmt}
                   \PY{p}{(}\PY{k}{do }\PY{p}{(}\PY{n+nb}{when }\PY{n+nv}{n} \PY{p}{(}\PY{n+nf}{eadd!} \PY{n+nv}{el} \PY{l+s+ss}{:cfNext} \PY{p}{(}\PY{n+nf}{cf\PYZhy{}peek} \PY{n+nv}{n}\PY{p}{)}\PY{p}{)}\PY{p}{)}
                       \PY{p}{(}\PY{n+nf}{recur} \PY{n+nv}{tail} \PY{n+nv}{exit} \PY{n+nv}{loop\PYZhy{}expr} \PY{n+nv}{loop\PYZhy{}succ} \PY{n+nv}{label\PYZhy{}succ\PYZhy{}map}\PY{p}{)}\PY{p}{)}
        \PY{l+s+ss}{\PYZsq{}flowgraph.Block}
                   \PY{p}{(}\PY{n+nf}{recur} \PY{p}{(}\PY{n+nb}{concat }\PY{p}{(}\PY{n+nf}{econtents} \PY{n+nv}{el}\PY{p}{)} \PY{n+nv}{tail}\PY{p}{)}
                          \PY{n+nv}{exit} \PY{n+nv}{loop\PYZhy{}expr} \PY{n+nv}{loop\PYZhy{}succ} \PY{n+nv}{label\PYZhy{}succ\PYZhy{}map}\PY{p}{)}
        \PY{l+s+ss}{\PYZsq{}flowgraph.Expr}
                   \PY{p}{(}\PY{k}{do }\PY{p}{(}\PY{n+nb}{when }\PY{n+nv}{n} \PY{p}{(}\PY{n+nf}{eadd!} \PY{n+nv}{el} \PY{l+s+ss}{:cfNext} \PY{p}{(}\PY{n+nf}{cf\PYZhy{}peek} \PY{n+nv}{n}\PY{p}{)}\PY{p}{)}\PY{p}{)}
                       \PY{p}{(}\PY{n+nf}{recur} \PY{n+nv}{tail} \PY{n+nv}{exit} \PY{n+nv}{loop\PYZhy{}expr} \PY{n+nv}{loop\PYZhy{}succ} \PY{n+nv}{label\PYZhy{}succ\PYZhy{}map}\PY{p}{)}\PY{p}{)}
        \PY{l+s+ss}{\PYZsq{}flowgraph.Label}
                   \PY{p}{(}\PY{n+nf}{recur} \PY{p}{(}\PY{n+nb}{cons }\PY{p}{(}\PY{n+nf}{eget} \PY{n+nv}{el} \PY{l+s+ss}{:stmt}\PY{p}{)} \PY{n+nv}{tail}\PY{p}{)} \PY{n+nv}{exit} \PY{n+nv}{loop\PYZhy{}expr} \PY{n+nv}{loop\PYZhy{}succ}
                          \PY{p}{(}\PY{n+nb}{assoc }\PY{n+nv}{label\PYZhy{}succ\PYZhy{}map} \PY{n+nv}{el} \PY{n+nv}{n}\PY{p}{)}\PY{p}{)}
        \PY{l+s+ss}{\PYZsq{}flowgraph.Return}
                   \PY{p}{(}\PY{k}{do }\PY{p}{(}\PY{n+nf}{eadd!} \PY{n+nv}{el} \PY{l+s+ss}{:cfNext} \PY{n+nv}{exit}\PY{p}{)}
                       \PY{p}{(}\PY{n+nf}{recur} \PY{n+nv}{tail} \PY{n+nv}{exit} \PY{n+nv}{loop\PYZhy{}expr} \PY{n+nv}{loop\PYZhy{}succ} \PY{n+nv}{label\PYZhy{}succ\PYZhy{}map}\PY{p}{)}\PY{p}{)}
        \PY{l+s+ss}{\PYZsq{}flowgraph.Break}
                   \PY{p}{(}\PY{k}{do }\PY{p}{(}\PY{n+nb}{if\PYZhy{}let }\PY{p}{[}\PY{n+nv}{l} \PY{p}{(}\PY{n+nf}{eget} \PY{n+nv}{el} \PY{l+s+ss}{:label}\PY{p}{)}\PY{p}{]}
                         \PY{p}{(}\PY{n+nf}{eadd!} \PY{n+nv}{el} \PY{l+s+ss}{:cfNext} \PY{p}{(}\PY{n+nf}{cf\PYZhy{}peek} \PY{p}{(}\PY{n+nf}{label\PYZhy{}succ\PYZhy{}map} \PY{n+nv}{l}\PY{p}{)}\PY{p}{)}\PY{p}{)}
                         \PY{p}{(}\PY{n+nf}{eadd!} \PY{n+nv}{el} \PY{l+s+ss}{:cfNext} \PY{p}{(}\PY{n+nf}{cf\PYZhy{}peek} \PY{n+nv}{loop\PYZhy{}succ}\PY{p}{)}\PY{p}{)}\PY{p}{)}
                       \PY{p}{(}\PY{n+nf}{recur} \PY{n+nv}{tail} \PY{n+nv}{exit} \PY{n+nv}{loop\PYZhy{}expr} \PY{n+nv}{loop\PYZhy{}succ} \PY{n+nv}{label\PYZhy{}succ\PYZhy{}map}\PY{p}{)}\PY{p}{)}
        \PY{l+s+ss}{\PYZsq{}flowgraph.Continue}
                   \PY{p}{(}\PY{k}{do }\PY{p}{(}\PY{n+nb}{if\PYZhy{}let }\PY{p}{[}\PY{n+nv}{l} \PY{p}{(}\PY{n+nf}{eget} \PY{n+nv}{el} \PY{l+s+ss}{:label}\PY{p}{)}\PY{p}{]}
                         \PY{p}{(}\PY{n+nf}{eadd!} \PY{n+nv}{el} \PY{l+s+ss}{:cfNext} \PY{p}{(}\PY{n+nf}{cf\PYZhy{}peek} \PY{n+nv}{l}\PY{p}{)}\PY{p}{)}
                         \PY{p}{(}\PY{n+nf}{eadd!} \PY{n+nv}{el} \PY{l+s+ss}{:cfNext} \PY{n+nv}{loop\PYZhy{}expr}\PY{p}{)}\PY{p}{)}
                       \PY{p}{(}\PY{n+nf}{recur} \PY{n+nv}{tail} \PY{n+nv}{exit} \PY{n+nv}{loop\PYZhy{}expr} \PY{n+nv}{loop\PYZhy{}succ} \PY{n+nv}{label\PYZhy{}succ\PYZhy{}map}\PY{p}{)}\PY{p}{)}
        \PY{l+s+ss}{\PYZsq{}flowgraph.Loop}
                   \PY{p}{(}\PY{k}{let }\PY{p}{[}\PY{p}{[}\PY{n+nv}{expr} \PY{n+nv}{body}\PY{p}{]} \PY{p}{(}\PY{n+nf}{econtents} \PY{n+nv}{el}\PY{p}{)}\PY{p}{]}
                     \PY{p}{(}\PY{n+nf}{recur} \PY{p}{(}\PY{n+nb}{cons }\PY{n+nv}{expr} \PY{p}{(}\PY{n+nb}{cons }\PY{n+nv}{body} \PY{p}{(}\PY{n+nb}{cons }\PY{n+nv}{expr} \PY{n+nv}{tail}\PY{p}{)}\PY{p}{)}\PY{p}{)}
                            \PY{n+nv}{exit} \PY{n+nv}{expr} \PY{n+nv}{n} \PY{n+nv}{label\PYZhy{}succ\PYZhy{}map}\PY{p}{)}\PY{p}{)}
        \PY{l+s+ss}{\PYZsq{}flowgraph.If}
                   \PY{p}{(}\PY{k}{let }\PY{p}{[}\PY{p}{[}\PY{n+nv}{expr} \PY{n+nv}{then} \PY{n+nv}{else}\PY{p}{]} \PY{p}{(}\PY{n+nf}{econtents} \PY{n+nv}{el}\PY{p}{)}\PY{p}{]}
                     \PY{p}{(}\PY{n+nf}{cf\PYZhy{}synth} \PY{p}{[}\PY{n+nv}{expr} \PY{n+nv}{then} \PY{p}{(}\PY{n+nf}{cf\PYZhy{}peek} \PY{n+nv}{n}\PY{p}{)}\PY{p}{]}
                               \PY{n+nv}{exit} \PY{n+nv}{loop\PYZhy{}expr} \PY{n+nv}{loop\PYZhy{}succ} \PY{n+nv}{label\PYZhy{}succ\PYZhy{}map}\PY{p}{)}
                     \PY{p}{(}\PY{k}{if }\PY{n+nv}{else}
                       \PY{p}{(}\PY{n+nf}{recur} \PY{p}{(}\PY{n+nb}{cons }\PY{n+nv}{expr} \PY{p}{(}\PY{n+nb}{cons }\PY{n+nv}{else} \PY{n+nv}{tail}\PY{p}{)}\PY{p}{)}
                              \PY{n+nv}{exit} \PY{n+nv}{loop\PYZhy{}expr} \PY{n+nv}{loop\PYZhy{}succ} \PY{n+nv}{label\PYZhy{}succ\PYZhy{}map}\PY{p}{)}
                       \PY{p}{(}\PY{n+nf}{recur} \PY{p}{(}\PY{n+nb}{cons }\PY{n+nv}{expr} \PY{n+nv}{tail}\PY{p}{)}
                              \PY{n+nv}{exit} \PY{n+nv}{loop\PYZhy{}expr} \PY{n+nv}{loop\PYZhy{}succ} \PY{n+nv}{label\PYZhy{}succ\PYZhy{}map}\PY{p}{)}\PY{p}{)}\PY{p}{)}
        \PY{l+s+ss}{\PYZsq{}flowgraph.Exit} \PY{p}{(}\PY{n+nb}{assert }\PY{p}{(}\PY{n+nb}{nil? }\PY{n+nv}{n}\PY{p}{)}\PY{p}{)}\PY{p}{)}\PY{p}{)}\PY{p}{)}\PY{p}{)}

\PY{p}{(}\PY{k+kd}{defn }\PY{n+nv}{synthesize\PYZhy{}cf\PYZhy{}edges} \PY{p}{[}\PY{n+nv}{model}\PY{p}{]}
  \PY{p}{(}\PY{n+nb}{doseq }\PY{p}{[}\PY{n+nv}{m} \PY{p}{(}\PY{n+nf}{eallobjects} \PY{n+nv}{model} \PY{l+s+ss}{\PYZsq{}flowgraph.Method}\PY{p}{)}
          \PY{l+s+ss}{:let} \PY{p}{[}\PY{n+nv}{exit} \PY{p}{(}\PY{n+nf}{the} \PY{p}{(}\PY{n+nf}{eallobjects} \PY{n+nv}{model} \PY{l+s+ss}{\PYZsq{}flowgraph.Exit}\PY{p}{)}\PY{p}{)}\PY{p}{]}\PY{p}{]}
    \PY{p}{(}\PY{n+nf}{cf\PYZhy{}synth} \PY{p}{[}\PY{n+nv}{m}\PY{p}{]} \PY{n+nv}{exit} \PY{n+nv}{nil} \PY{n+nv}{nil} \PY{n+nv}{nil}\PY{p}{)}\PY{p}{)}\PY{p}{)}
\end{Verbatim}

\section{The complete Data Flow Transformation}
\label{sec:complete-data-flow}

\begin{Verbatim}[commandchars=\\\{\},fontsize=\footnotesize]
\PY{p}{(}\PY{k+kd}{defn }\PY{n+nv}{find\PYZhy{}nearest\PYZhy{}definers} \PY{p}{[}\PY{n+nv}{fi} \PY{n+nv}{uv}\PY{p}{]}
  \PY{p}{(}\PY{k}{loop }\PY{p}{[}\PY{n+nv}{preds} \PY{p}{(}\PY{n+nb}{mapcat }\PY{o}{\PYZsh{}}\PY{p}{(}\PY{n+nf}{adjs} \PY{n+nv}{\PYZpc{}} \PY{l+s+ss}{:cfPrev}\PY{p}{)} \PY{p}{(}\PY{k}{if }\PY{p}{(}\PY{n+nf}{coll?} \PY{n+nv}{fi}\PY{p}{)} \PY{n+nv}{fi} \PY{p}{[}\PY{n+nv}{fi}\PY{p}{]}\PY{p}{)}\PY{p}{)}
         \PY{n+nv}{r} \PY{p}{[}\PY{p}{]}
         \PY{n+nv}{known} \PY{o}{\PYZsh{}}\PY{p}{\PYZob{}}\PY{p}{\PYZcb{}}\PY{p}{]}
    \PY{p}{(}\PY{k}{if }\PY{p}{(}\PY{n+nb}{seq }\PY{n+nv}{preds}\PY{p}{)}
      \PY{p}{(}\PY{k}{let }\PY{p}{[}\PY{n+nv}{definers} \PY{p}{(}\PY{n+nb}{filter }\PY{o}{\PYZsh{}}\PY{p}{(}\PY{n+nf}{member?} \PY{n+nv}{uv} \PY{p}{(}\PY{n+nf}{eget} \PY{n+nv}{\PYZpc{}} \PY{l+s+ss}{:def}\PY{p}{)}\PY{p}{)} \PY{n+nv}{preds}\PY{p}{)}
            \PY{n+nv}{others}   \PY{p}{(}\PY{n+nb}{remove }\PY{o}{\PYZsh{}}\PY{p}{(}\PY{n+nf}{member?} \PY{n+nv}{uv} \PY{p}{(}\PY{n+nf}{eget} \PY{n+nv}{\PYZpc{}} \PY{l+s+ss}{:def}\PY{p}{)}\PY{p}{)} \PY{n+nv}{preds}\PY{p}{)}\PY{p}{]}
        \PY{p}{(}\PY{n+nf}{recur} \PY{p}{(}\PY{n+nb}{remove }\PY{o}{\PYZsh{}}\PY{p}{(}\PY{n+nf}{member?} \PY{n+nv}{\PYZpc{}} \PY{n+nv}{known}\PY{p}{)} \PY{p}{(}\PY{n+nb}{mapcat }\PY{o}{\PYZsh{}}\PY{p}{(}\PY{n+nf}{adjs} \PY{n+nv}{\PYZpc{}} \PY{l+s+ss}{:cfPrev}\PY{p}{)} \PY{n+nv}{others}\PY{p}{)}\PY{p}{)}
               \PY{p}{(}\PY{n+nb}{into }\PY{n+nv}{r} \PY{n+nv}{definers}\PY{p}{)}
               \PY{p}{(}\PY{n+nb}{into }\PY{n+nv}{known} \PY{n+nv}{preds}\PY{p}{)}\PY{p}{)}\PY{p}{)}
      \PY{n+nv}{r}\PY{p}{)}\PY{p}{)}\PY{p}{)}

\PY{p}{(}\PY{k+kd}{defn }\PY{n+nv}{synthesize\PYZhy{}df\PYZhy{}edges} \PY{p}{[}\PY{n+nv}{model}\PY{p}{]}
  \PY{p}{(}\PY{n+nb}{doseq }\PY{p}{[}\PY{n+nv}{fi} \PY{p}{(}\PY{n+nf}{eallobjects} \PY{n+nv}{model} \PY{l+s+ss}{\PYZsq{}flowgraph.FlowInstr}\PY{p}{)}
          \PY{n+nv}{used\PYZhy{}var} \PY{p}{(}\PY{n+nf}{eget} \PY{n+nv}{fi} \PY{l+s+ss}{:use}\PY{p}{)}
          \PY{n+nv}{nearest\PYZhy{}definer} \PY{p}{(}\PY{n+nf}{find\PYZhy{}nearest\PYZhy{}definers} \PY{n+nv}{fi} \PY{n+nv}{used\PYZhy{}var}\PY{p}{)}\PY{p}{]}
    \PY{p}{(}\PY{n+nf}{eadd!} \PY{n+nv}{nearest\PYZhy{}definer} \PY{l+s+ss}{:dfNext} \PY{n+nv}{fi}\PY{p}{)}\PY{p}{)}
  \PY{p}{(}\PY{n+nb}{doseq }\PY{p}{[}\PY{n+nv}{v} \PY{p}{(}\PY{n+nf}{vec} \PY{p}{(}\PY{n+nf}{eallobjects} \PY{n+nv}{model} \PY{l+s+ss}{\PYZsq{}Var}\PY{p}{)}\PY{p}{)}\PY{p}{]}
    \PY{p}{(}\PY{n+nf}{edelete!} \PY{n+nv}{v}\PY{p}{)}\PY{p}{)}\PY{p}{)}
\end{Verbatim}

\section{The complete Validation DSL Implementation}
\label{sec:compl-valid-dsl}

\begin{Verbatim}[commandchars=\\\{\},fontsize=\footnotesize]
\PY{p}{(}\PY{k+kd}{defn }\PY{n+nv}{run\PYZhy{}flowgraph\PYZhy{}transformations} \PY{p}{[}\PY{n+nv}{file}\PY{p}{]}
  \PY{p}{(}\PY{n+nf}{System/gc}\PY{p}{)}
  \PY{p}{(}\PY{n+nb}{println }\PY{l+s}{\PYZdq{}Running Transformation on\PYZdq{}} \PY{n+nv}{file}\PY{p}{)}
  \PY{p}{(}\PY{n+nb}{print }\PY{l+s}{\PYZdq{}Load Time: \PYZdq{}}\PY{p}{)}
  \PY{p}{(}\PY{k}{let }\PY{p}{[}\PY{n+nv}{jamopp\PYZhy{}model} \PY{p}{(}\PY{n+nb}{time }\PY{p}{(}\PY{n+nf}{load\PYZhy{}model} \PY{n+nv}{file}\PY{p}{)}\PY{p}{)}
        \PY{n+nv}{outfile} \PY{p}{(}\PY{n+nf}{str/replace} \PY{p}{(}\PY{n+nf}{str/replace} \PY{n+nv}{file} \PY{l+s}{\PYZdq{}models/\PYZdq{}} \PY{l+s}{\PYZdq{}results/\PYZdq{}}\PY{p}{)}
                             \PY{l+s}{\PYZdq{}.java.xmi\PYZdq{}} \PY{l+s}{\PYZdq{}.xmi\PYZdq{}}\PY{p}{)}
        \PY{n+nv}{outvizfile} \PY{p}{(}\PY{n+nf}{str/replace} \PY{n+nv}{outfile} \PY{l+s}{\PYZdq{}.xmi\PYZdq{}} \PY{l+s}{\PYZdq{}.pdf\PYZdq{}}\PY{p}{)}
        \PY{n+nv}{fg\PYZhy{}trg} \PY{p}{(}\PY{n+nf}{new\PYZhy{}model}\PY{p}{)}\PY{p}{]}
    \PY{p}{(}\PY{n+nb}{println }\PY{l+s}{\PYZdq{}Execution Times:\PYZdq{}}\PY{p}{)}
    \PY{p}{(}\PY{n+nb}{print }\PY{l+s}{\PYZdq{}  \PYZhy{} JaMoPP to StructureGraph (with Vars): \PYZdq{}}\PY{p}{)}
    \PY{p}{(}\PY{n+nb}{time }\PY{p}{(}\PY{n+nf}{java2flowgraph} \PY{n+nv}{jamopp\PYZhy{}model} \PY{n+nv}{fg\PYZhy{}trg}\PY{p}{)}\PY{p}{)}
    \PY{p}{(}\PY{n+nb}{print }\PY{l+s}{\PYZdq{}  \PYZhy{} Control Flow Analysis:                \PYZdq{}}\PY{p}{)}
    \PY{p}{(}\PY{n+nb}{time }\PY{p}{(}\PY{n+nf}{synthesize\PYZhy{}cf\PYZhy{}edges} \PY{n+nv}{fg\PYZhy{}trg}\PY{p}{)}\PY{p}{)}
    \PY{p}{(}\PY{n+nb}{print }\PY{l+s}{\PYZdq{}  \PYZhy{} Data Flow Analysis:                   \PYZdq{}}\PY{p}{)}
    \PY{p}{(}\PY{n+nb}{time }\PY{p}{(}\PY{n+nf}{synthesize\PYZhy{}df\PYZhy{}edges} \PY{n+nv}{fg\PYZhy{}trg}\PY{p}{)}\PY{p}{)}
    \PY{p}{(}\PY{n+nf}{save\PYZhy{}model} \PY{n+nv}{fg\PYZhy{}trg} \PY{n+nv}{outfile}\PY{p}{)}
    \PY{p}{(}\PY{n+nb}{when }\PY{p}{(}\PY{n+nb}{\PYZlt{} }\PY{p}{(}\PY{n+nb}{count }\PY{p}{(}\PY{n+nf}{eallobjects} \PY{n+nv}{fg\PYZhy{}trg}\PY{p}{)}\PY{p}{)} \PY{l+m+mi}{80}\PY{p}{)}
      \PY{p}{(}\PY{n+nf}{print\PYZhy{}model} \PY{n+nv}{fg\PYZhy{}trg} \PY{n+nv}{outvizfile}\PY{p}{)}\PY{p}{)}
    \PY{n+nv}{fg\PYZhy{}trg}\PY{p}{)}\PY{p}{)}

\PY{p}{(}\PY{k+kd}{defmacro }\PY{n+nv}{make\PYZhy{}test} \PY{p}{[}\PY{n+nv}{n} \PY{n+nv}{file} \PY{n+nv}{expected\PYZhy{}cfs} \PY{n+nv}{expected\PYZhy{}dfs}\PY{p}{]}
  \PY{o}{`}\PY{p}{(}\PY{n+nf}{deftest} \PY{o}{\PYZti{}}\PY{n+nv}{n}
     \PY{p}{(}\PY{n+nb}{println }\PY{l+s}{\PYZdq{}========================================================================\PYZdq{}}\PY{p}{)}
     \PY{p}{(}\PY{k}{let }\PY{p}{[}\PY{n+nv}{fg\PYZhy{}trg\PYZsh{}} \PY{p}{(}\PY{n+nf}{run\PYZhy{}flowgraph\PYZhy{}transformations} \PY{o}{\PYZti{}}\PY{n+nv}{file}\PY{p}{)}
           \PY{n+nv}{exp\PYZhy{}cfs\PYZsh{}} \PY{o}{\PYZti{}}\PY{n+nv}{expected\PYZhy{}cfs}
           \PY{n+nv}{exp\PYZhy{}dfs\PYZsh{}} \PY{o}{\PYZti{}}\PY{n+nv}{expected\PYZhy{}dfs}
           \PY{n+nv}{cfs\PYZsh{}} \PY{p}{(}\PY{n+nb}{set }\PY{p}{(}\PY{n+nb}{map }\PY{p}{(}\PY{k}{fn }\PY{p}{[}\PY{p}{[}\PY{n+nv}{s\PYZsh{}} \PY{n+nv}{t\PYZsh{}}\PY{p}{]}\PY{p}{]} \PY{p}{[}\PY{p}{(}\PY{n+nf}{eget} \PY{n+nv}{s\PYZsh{}} \PY{l+s+ss}{:txt}\PY{p}{)} \PY{p}{(}\PY{n+nf}{eget} \PY{n+nv}{t\PYZsh{}} \PY{l+s+ss}{:txt}\PY{p}{)}\PY{p}{]}\PY{p}{)}
                          \PY{p}{(}\PY{n+nf}{ecrosspairs} \PY{n+nv}{fg\PYZhy{}trg\PYZsh{}} \PY{l+s+ss}{:cfPrev} \PY{l+s+ss}{:cfNext}\PY{p}{)}\PY{p}{)}\PY{p}{)}
           \PY{n+nv}{dfs\PYZsh{}} \PY{p}{(}\PY{n+nb}{set }\PY{p}{(}\PY{n+nb}{map }\PY{p}{(}\PY{k}{fn }\PY{p}{[}\PY{p}{[}\PY{n+nv}{s\PYZsh{}} \PY{n+nv}{t\PYZsh{}}\PY{p}{]}\PY{p}{]} \PY{p}{[}\PY{p}{(}\PY{n+nf}{eget} \PY{n+nv}{s\PYZsh{}} \PY{l+s+ss}{:txt}\PY{p}{)} \PY{p}{(}\PY{n+nf}{eget} \PY{n+nv}{t\PYZsh{}} \PY{l+s+ss}{:txt}\PY{p}{)}\PY{p}{]}\PY{p}{)}
                          \PY{p}{(}\PY{n+nf}{ecrosspairs} \PY{n+nv}{fg\PYZhy{}trg\PYZsh{}} \PY{n+nv}{nil} \PY{l+s+ss}{:dfNext}\PY{p}{)}\PY{p}{)}\PY{p}{)}\PY{p}{]}
       \PY{p}{(}\PY{n+nf}{cond}
        \PY{p}{(}\PY{n+nf}{set?} \PY{n+nv}{exp\PYZhy{}cfs\PYZsh{}}\PY{p}{)} \PY{p}{(}\PY{k}{let }\PY{p}{[}\PY{n+nv}{cf\PYZhy{}d1\PYZsh{}} \PY{p}{(}\PY{n+nf}{clojure.set/difference} \PY{n+nv}{exp\PYZhy{}cfs\PYZsh{}} \PY{n+nv}{cfs\PYZsh{}}\PY{p}{)}
                              \PY{n+nv}{cf\PYZhy{}d2\PYZsh{}} \PY{p}{(}\PY{n+nf}{clojure.set/difference} \PY{n+nv}{cfs\PYZsh{}} \PY{n+nv}{exp\PYZhy{}cfs\PYZsh{}}\PY{p}{)}\PY{p}{]}
                          \PY{p}{(}\PY{n+nf}{is} \PY{p}{(}\PY{n+nf}{empty?} \PY{n+nv}{cf\PYZhy{}d1\PYZsh{}}\PY{p}{)} \PY{l+s}{\PYZdq{}Missing cf\PYZhy{}edges\PYZdq{}}\PY{p}{)}
                          \PY{p}{(}\PY{n+nf}{is} \PY{p}{(}\PY{n+nf}{empty?} \PY{n+nv}{cf\PYZhy{}d2\PYZsh{}}\PY{p}{)} \PY{l+s}{\PYZdq{}Too many cf\PYZhy{}edges\PYZdq{}}\PY{p}{)}\PY{p}{)}
        \PY{p}{(}\PY{n+nf}{number?} \PY{n+nv}{exp\PYZhy{}cfs\PYZsh{}}\PY{p}{)} \PY{p}{(}\PY{n+nf}{do}
                             \PY{p}{(}\PY{n+nb}{println }\PY{l+s}{\PYZdq{}Only checking number of cfNext links.\PYZdq{}}\PY{p}{)}
                             \PY{p}{(}\PY{n+nf}{is} \PY{p}{(}\PY{n+nb}{= }\PY{n+nv}{exp\PYZhy{}cfs\PYZsh{}} \PY{p}{(}\PY{n+nb}{count }\PY{n+nv}{cfs\PYZsh{}}\PY{p}{)}\PY{p}{)}\PY{p}{)}\PY{p}{)}
        \PY{l+s+ss}{:else} \PY{p}{(}\PY{n+nb}{println }\PY{l+s}{\PYZdq{}No expected cfNext links given.\PYZdq{}}\PY{p}{)}\PY{p}{)}
       \PY{p}{(}\PY{n+nf}{cond}
        \PY{p}{(}\PY{n+nf}{set?} \PY{n+nv}{exp\PYZhy{}dfs\PYZsh{}}\PY{p}{)} \PY{p}{(}\PY{k}{let }\PY{p}{[}\PY{n+nv}{df\PYZhy{}d1\PYZsh{}} \PY{p}{(}\PY{n+nf}{clojure.set/difference} \PY{n+nv}{exp\PYZhy{}dfs\PYZsh{}} \PY{n+nv}{dfs\PYZsh{}}\PY{p}{)}
                              \PY{n+nv}{df\PYZhy{}d2\PYZsh{}} \PY{p}{(}\PY{n+nf}{clojure.set/difference} \PY{n+nv}{dfs\PYZsh{}} \PY{n+nv}{exp\PYZhy{}dfs\PYZsh{}}\PY{p}{)}\PY{p}{]}
                          \PY{p}{(}\PY{n+nf}{is} \PY{p}{(}\PY{n+nf}{empty?} \PY{n+nv}{df\PYZhy{}d1\PYZsh{}}\PY{p}{)} \PY{l+s}{\PYZdq{}Missing df\PYZhy{}edges\PYZdq{}}\PY{p}{)}
                          \PY{p}{(}\PY{n+nf}{is} \PY{p}{(}\PY{n+nf}{empty?} \PY{n+nv}{df\PYZhy{}d2\PYZsh{}}\PY{p}{)} \PY{l+s}{\PYZdq{}Too many df\PYZhy{}edges\PYZdq{}}\PY{p}{)}\PY{p}{)}
        \PY{p}{(}\PY{n+nf}{number?} \PY{n+nv}{exp\PYZhy{}dfs\PYZsh{}}\PY{p}{)} \PY{p}{(}\PY{n+nf}{do}
                             \PY{p}{(}\PY{n+nb}{println }\PY{l+s}{\PYZdq{}Only checking number of dfNext links.\PYZdq{}}\PY{p}{)}
                             \PY{p}{(}\PY{n+nf}{is} \PY{p}{(}\PY{n+nb}{= }\PY{n+nv}{exp\PYZhy{}dfs\PYZsh{}} \PY{p}{(}\PY{n+nb}{count }\PY{n+nv}{dfs\PYZsh{}}\PY{p}{)}\PY{p}{)}\PY{p}{)}\PY{p}{)}
        \PY{l+s+ss}{:else} \PY{p}{(}\PY{n+nb}{println }\PY{l+s}{\PYZdq{}No expected dfNext links given.\PYZdq{}}\PY{p}{)}\PY{p}{)}\PY{p}{)}\PY{p}{)}\PY{p}{)}
\end{Verbatim}

\subsection{Two Example Validation Specifications}
\label{sec:two-example-valid}

\begin{Verbatim}[commandchars=\\\{\},fontsize=\footnotesize]
\PY{p}{(}\PY{n+nf}{make\PYZhy{}test} \PY{n+nv}{test\PYZhy{}fg\PYZhy{}transform\PYZhy{}test4} \PY{l+s}{\PYZdq{}models/Test4.java.xmi\PYZdq{}}
           \PY{o}{\PYZsh{}}\PY{p}{\PYZob{}}\PY{p}{[}\PY{l+s}{\PYZdq{}testMethod()\PYZdq{}} \PY{l+s}{\PYZdq{}int i = 100;\PYZdq{}}\PY{p}{]}
             \PY{p}{[}\PY{l+s}{\PYZdq{}int i = 100;\PYZdq{}} \PY{l+s}{\PYZdq{}i \PYZgt{} 0\PYZdq{}}\PY{p}{]}
             \PY{p}{[}\PY{l+s}{\PYZdq{}i \PYZgt{} 0\PYZdq{}}        \PY{l+s}{\PYZdq{}Exit\PYZdq{}}\PY{p}{]}
             \PY{p}{[}\PY{l+s}{\PYZdq{}i \PYZgt{} 0\PYZdq{}}        \PY{l+s}{\PYZdq{}i \PYZgt{} 50\PYZdq{}}\PY{p}{]}
             \PY{p}{[}\PY{l+s}{\PYZdq{}i \PYZgt{} 50\PYZdq{}}       \PY{l+s}{\PYZdq{}i\PYZhy{}\PYZhy{};\PYZdq{}}\PY{p}{]}
             \PY{p}{[}\PY{l+s}{\PYZdq{}i \PYZgt{} 50\PYZdq{}}       \PY{l+s}{\PYZdq{}i = i \PYZhy{} 10;\PYZdq{}}\PY{p}{]}
             \PY{p}{[}\PY{l+s}{\PYZdq{}i = i \PYZhy{} 10;\PYZdq{}}  \PY{l+s}{\PYZdq{}i == 50\PYZdq{}}\PY{p}{]}
             \PY{p}{[}\PY{l+s}{\PYZdq{}i == 50\PYZdq{}}      \PY{l+s}{\PYZdq{}break;\PYZdq{}}\PY{p}{]}
             \PY{p}{[}\PY{l+s}{\PYZdq{}i == 50\PYZdq{}}      \PY{l+s}{\PYZdq{}i \PYZgt{} 50\PYZdq{}}\PY{p}{]}
             \PY{p}{[}\PY{l+s}{\PYZdq{}break;\PYZdq{}}       \PY{l+s}{\PYZdq{}i\PYZhy{}\PYZhy{};\PYZdq{}}\PY{p}{]}
             \PY{p}{[}\PY{l+s}{\PYZdq{}i\PYZhy{}\PYZhy{};\PYZdq{}}         \PY{l+s}{\PYZdq{}i \PYZgt{} 0\PYZdq{}}\PY{p}{]}\PY{p}{\PYZcb{}}
           \PY{o}{\PYZsh{}}\PY{p}{\PYZob{}}\PY{p}{[}\PY{l+s}{\PYZdq{}int i = 100;\PYZdq{}} \PY{l+s}{\PYZdq{}i \PYZgt{} 0\PYZdq{}}\PY{p}{]}
             \PY{p}{[}\PY{l+s}{\PYZdq{}int i = 100;\PYZdq{}} \PY{l+s}{\PYZdq{}i \PYZgt{} 50\PYZdq{}}\PY{p}{]}
             \PY{p}{[}\PY{l+s}{\PYZdq{}int i = 100;\PYZdq{}} \PY{l+s}{\PYZdq{}i = i \PYZhy{} 10;\PYZdq{}}\PY{p}{]}
             \PY{p}{[}\PY{l+s}{\PYZdq{}int i = 100;\PYZdq{}} \PY{l+s}{\PYZdq{}i\PYZhy{}\PYZhy{};\PYZdq{}}\PY{p}{]}
             \PY{p}{[}\PY{l+s}{\PYZdq{}i = i \PYZhy{} 10;\PYZdq{}}  \PY{l+s}{\PYZdq{}i == 50\PYZdq{}}\PY{p}{]}
             \PY{p}{[}\PY{l+s}{\PYZdq{}i = i \PYZhy{} 10;\PYZdq{}}  \PY{l+s}{\PYZdq{}i \PYZgt{} 50\PYZdq{}}\PY{p}{]}
             \PY{p}{[}\PY{l+s}{\PYZdq{}i = i \PYZhy{} 10;\PYZdq{}}  \PY{l+s}{\PYZdq{}i = i \PYZhy{} 10;\PYZdq{}}\PY{p}{]}
             \PY{p}{[}\PY{l+s}{\PYZdq{}i = i \PYZhy{} 10;\PYZdq{}}  \PY{l+s}{\PYZdq{}i\PYZhy{}\PYZhy{};\PYZdq{}}\PY{p}{]}
             \PY{p}{[}\PY{l+s}{\PYZdq{}i\PYZhy{}\PYZhy{};\PYZdq{}}         \PY{l+s}{\PYZdq{}i \PYZgt{} 0\PYZdq{}}\PY{p}{]}
             \PY{p}{[}\PY{l+s}{\PYZdq{}i\PYZhy{}\PYZhy{};\PYZdq{}}         \PY{l+s}{\PYZdq{}i \PYZgt{} 50\PYZdq{}}\PY{p}{]}
             \PY{p}{[}\PY{l+s}{\PYZdq{}i\PYZhy{}\PYZhy{};\PYZdq{}}         \PY{l+s}{\PYZdq{}i = i \PYZhy{} 10;\PYZdq{}}\PY{p}{]}
             \PY{p}{[}\PY{l+s}{\PYZdq{}i\PYZhy{}\PYZhy{};\PYZdq{}}         \PY{l+s}{\PYZdq{}i\PYZhy{}\PYZhy{};\PYZdq{}}\PY{p}{]}\PY{p}{\PYZcb{}}\PY{p}{)}

\PY{c+c1}{;; For the large models, only the correct number of cfNext/dfNext links is asserted.}
\PY{p}{(}\PY{n+nf}{make\PYZhy{}test} \PY{n+nv}{test\PYZhy{}fg\PYZhy{}transform\PYZhy{}test9} \PY{l+s}{\PYZdq{}models/Test9.java.xmi\PYZdq{}} \PY{l+m+mi}{14452} \PY{l+m+mi}{27202}\PY{p}{)}
\end{Verbatim}

\end{document}